\documentclass[12pt,letter]{article}
\usepackage[left=1in,right=1in,top=1in,bottom=1in]{geometry}
\usepackage{amsmath}
\usepackage{amssymb}
\usepackage{graphics}
\usepackage{rotating}
\usepackage[dvips]{epsfig}
\usepackage{natbib}
\usepackage[ansinew]{inputenc} % damit Umlaute direkt eingegeben werden kï¿½nnen

\usepackage{amsfonts}
\usepackage{amsbsy}
\usepackage{rotating}
\usepackage[flushleft]{threeparttable}
\usepackage[margin=10pt,font=small,labelfont=bf,labelsep=period]{caption}

\usepackage{adjustbox}		% to resize tables to textwidth
\usepackage{booktabs}
\usepackage{multirow}

\usepackage{subcaption}		% to create subfigures
\usepackage{graphicx} 

\usepackage{setspace}
\newcommand{\ba}{\begin{array}}
	\newcommand{\ea}{\end{array}}

\newcommand{\be}{\begin{equation*}}
\newcommand{\ee}{\end{equation*}}
\newcommand{\bea}{\begin{eqnarray*}}
	\newcommand{\eea}{\end{eqnarray*}}

\newcommand{\var}{\mbox{Var}}

\newcommand{\E}{\mbox{E}}

\newtheorem{proposition}{Proposition}

\newdimen\dummy
\dummy=\oddsidemargin
\begin{document}

%\title{Forecasting under Long Memory\thanks{\textbf{Acknowledgements:}Earlier versions of this paper were presented ... The authors thank .}}
%
%\author{Marc-Oliver Pohle and Uwe Hassler\thanks{Corresponding author: hassler@wiwi.uni-frankfurt.de}\\
%
% {Goethe University Frankfurt\thanks{Campus Westend, Economics and Business Administration, 
%		60323 Frankfurt, Germany}}}
%\date{ \today }
%\maketitle

\title{Forecasting under Long Memory and Nonstationarity\thanks{
		\textbf{Acknowledgements:} We thank Richard T.\ Baillie, Federico M.\ Bandi, A.\ Ronald Galant, Patrik Guggenberger, Mehdi Hosseinkouchack, Joon Y.\ Park and Antonio Rubia for many helpful comments. Earlier versions of this paper were presented at Indiana University, Bloomington, at Penn State University, the 2nd Conference on New Trends and Developments in Econometrics (Bank of Portugal), the Statistische Woche 2018 in Linz, the Hannover Workshop on Long Memory in October 2018 and the 
		International Association for Applied Econometrics Annual Conference 2019 in Nicosia.}}
	
\author{Uwe Hassler and Marc-Oliver Pohle\thanks{Corresponding Author: pohle@econ.uni-frankfurt.de, Address: Campus Westend, RuW
		Building, 60323 Frankfurt, Germany} \\
	Goethe University Frankfurt}	 

\maketitle

\begin{abstract}
	
Long memory in the sense of slowly decaying autocorrelations is a stylized fact in many time series from economics and finance. The fractionally integrated process is the workhorse model for the analysis of these time series. Nevertheless, there is mixed evidence in the literature concerning its usefulness for forecasting and how forecasting based on it should be implemented. 

Employing pseudo-out-of-sample forecasting on inflation and realized volatility time series and simulations we show that methods based on fractional integration clearly are superior to alternative methods not accounting for long memory, including autoregressions and exponential smoothing. Our proposal of choosing a fixed fractional integration parameter of $d=0.5$ a priori yields the best results overall, capturing long memory behavior, but overcoming the deficiencies of methods using an estimated parameter.

Regarding the implementation of forecasting methods based on fractional integration, we use simulations to compare local and global semiparametric and parametric estimators of the long memory parameter from the Whittle family and provide asymptotic theory backed up by simulations to compare different mean estimators. Both of these analyses lead to new results, which are also of interest outside the realm of forecasting.

%\textbf{MSC:} 62F03 (hypothesis testing), 62F10 (point estimation), 62M10 (time series)
\end{abstract}

\section{Introduction}

Long memory or strong persistence are considered as stylized facts in many time series from economics and finance. The strong persistence characterized by slowly decaying autocorrelations on the one hand makes modeling and subsequent estimation hard for these series. Fractional integration [FI] of order $d$ is the most widely used model to capture strong persistence and long memory. Estimated values of $d$ vary between 0 and 1, with $d=1/2$ separating the reign of stationarity from nonstationarity.  FI thus  offers a lot of  flexibility in modeling persistence. This is a virtue  and a burden at the same time since the estimation of $d$ is notoriously difficult and troubled by large variances of slowly converging semiparametric estimators. On the other hand, the strong persistence in these series makes promises with regard to their forecastability. The past of long memory time series contains more information about their future compared to less persistent series, which can be exploited to produce high quality forecasts even for quite long forecasting horizons if the above-mentioned problems can be overcome.\footnote{Note, however, that we are interested in weaker persistence than that generated by an autoregressive root close to one; for forecasting in such a trending environment see e.g.\ \citet[chapter 20]{Elliott2016}.} 

Driven by the empirical relevance and the theoretical challenges in modeling and estimation a large literature on long memory time series and FI has emerged since the seminal work by \citet{GrangerJoyeux80} and \citet{GPH83}, with one major motivation being forecasting. Nevertheless, there is a lot of mixed evidence in the literature focusing explicitly on forecasting under long memory and many open questions remain, which we try to shed light on in this paper. Before taking a look at these unresolved issues, we shortly review the related literature.

Early simulation-based evidence on forecasting under long memory is provided by \citet{Ray1993} and \citet{CratoRay96}. They examine the usefulness of stationary autoregressive fractionally integrated moving average [ARFIMA] models for forecasting compared to autoregressions of large order, so-called long autoregressions [LARs]. In their simulations, they find inferior or at most equal predictive performance of the ARFIMA models, which they attribute to the difficulties in model selection and estimation of the long memory parameter $d$, and thus question the usefulness of these models for prediction. \cite{SmithYadav94} analyze the effects of using ARIMA($p$,1,$q$) models for forecasting nonstationary ARFIMA models with $d$ ranging from 0.7 to 1.3. They conclude that the losses from overdifferencing are minor, even though they do not account for the estimation uncertainty from the estimation of $d$, while underdifferencing can lead to more substantial losses. \cite{BrodskyHurvich99} demonstrate that FI models are superior in forecasting long memory time series compared to ARMA(1,1) models. \cite{BhardwajSwanson06} report gains in predictive performance when using FI models compared to a number of short memory competitors for medium sample sizes. For larger sample sizes, these gains are reported to grow substantially.

Concerning empirical evidence, again \cite{BhardwajSwanson06} report predictive superiority of FI models for financial data, but less clear-cut evidence for macroeconomic data. They attribute this finding mainly to the much larger sample sizes in finance, which lead to a more precise estimation of the long memory parameter. Furthermore, since \cite{Andersenetal03} FI models are widely used in realized volatility modeling and forecasting, see e.g.\ the special issue in Econometric Reviews edited by \cite{MaasoumiMcAleer08} or \cite{Proietti16} for a more recent example, and several papers analyze their performance compared to short memory competitors, see \cite{HansenLunde2011} for an overview. The evidence is mixed as some authors claim equal predictive ability of short memory alternatives like the Heterogeneous Autoregressive model [HAR] by \cite{corsi2009}; see also \cite{Baillie19} for recent evidence on the roles of the FI and the HAR model for modeling realized volatility.

Given the state of the literature on forecasting under long memory, several open issues remain: As the short review shows, a central question around which the literature has been circling in the past 20 years by providing evidence in both directions is the following: Does using long memory models when forecasting time series that exhibit strong persistence improve predictive performance or do simple short memory alternatives perform at least as well? Our first contribution lies in providing new evidence to answer this question: In systematic forecasting experiments using classical examples of strongly persistent time series from macroeconomics and finance, namely inflation and realized volatility, as well as simulated series, we find that FI models outperform a wide range of competing methods, including autoregressive [AR] models, exponential smoothing [ES] methods and for the realized volatility series also the HAR model. We come up with the idea of using a fixed-$d$ FI model with $d=0.5$, which almost uniformly outperforms all non-FI competitors and, even for large sample sizes, shows a comparable performance to FI models with estimated $d$. Under certain practically relevant circumstances like small estimation samples and more persistent short memory components, we find substantial gains of the fixed-$d$ method, which increase with the forecasting horizon. Picking $d=0.5$ is a natural choice at the border of nonstationarity and somehow in the middle between the classical $I(0)$ and $I(1)$ paradigms, robustifying the forecasting process against strong persistence and not being troubled with the estimation of the exact value of $d$. These surprising results do not only equip us with a simple and successful method for forecasting under long memory, but also provide an explanation for the mixed evidence formerly reported in the literature: Our analysis suggests that when the task is to forecast a strongly persistent time series, the inclusion of a long memory component into the used model is beneficial under all examined circumstances, e.g.\ over all sample sizes and forecasting horizons. However, FI models with estimated $d$ may sometimes be outperformed by AR models in small and medium-sized samples as parts of the literature report. This may be caused by a bad specification of the forecasting method with e.g.\ an unfavorable estimator of the long memory parameter $d$ or an inappropriate bandwidth choice for an otherwise well-suited semiparametric estimator. But even if a sensible specification is used this will happen due the fundamentally high uncertainty plaguing the estimation of $d$.

As our second contribution we analyze how to exactly construct good forecasts using FI models. Naturally, this becomes relevant after having provided evidence that FI models are useful in forecasting strongly persistent time series, but of course has to be answered in the first place to even establish this evidence. We focus on three crucial aspects here. The first one is the estimation of $d$. Because of the inherent difficulty of this task, there is a large literature on it and an ongoing debate, by which applied researchers or forecasting professionals can easily be overwhelmed. We carry out a simulation study to compare parametric, local and global semiparametric estimation methods, focusing on the representatives from the Whittle family of estimators since they have favorable theoretical properties and show a good practical performance in general. For the parametric Whittle estimator we use Akaike's information criterion [AIC] to select a model and make the estimator feasible. We find quite dramatic losses of the parametric and the global semiparametric estimators compared to the local Whittle, which we can trace back to the bad performance of the Whittle estimator even under small amounts of misspecification. Consequently, we use the local Whittle estimator with different ad hoc bandwidth choices for our forecasting experiments. The second aspect is the estimation of the mean, which is usually not even mentioned in the long  memory forecasting literature, even though it is well known that this is a hard task under strong persistence. We compare the arithmetic mean to the estimators proposed by \cite{Robinson94EffTests} and \cite{Shimotsu10} through a simulation study and find that the estimator of Robinson performs best. To the best of our knowledge, the theoretical properties of Robinson's estimator have not been analyzed before. We provide asymptotic theory, which shows that substantial efficiency gains can be made, explaining our simulation results. The third aspect concerns the following question: How much past information should be optimally used for estimating the model used for forecasting? While earlier work as e.g.\ \cite{BhardwajSwanson06} implicitly or explicitly assumes that the complete available past of the series of interest should be used because of the decreased estimation uncertainty, we find that this is not necessarily the case (see also \cite{pesaran2007}): Our empirical results suggest that a medium-sized rolling forecasting window seems to yield the best forecasting performance, balancing out decreased estimation uncertainty and ongoing gradual structural change optimally.

As outlined in the previous paragraph, the third contribution of this paper is to provide new results of interest also outside the realm of forecasting concerning the estimation of the long memory parameter and the mean under long memory.

The rest of the paper is structured as follows: In the next section we review the classical approach to modeling strong persistence, namely fractional integration. In the third section, the estimation of the long memory parameter is treated, while the fourth section deals with the estimation of the mean. In the fifth section, we describe the forecasting methods, i.e.\ on the one hand how we specifically implement forecasting with FI models based on the evidence from the previous two sections, and on the other hand what the competing methods are and why we use them. The sixth and the seventh section contain the results from the empirical and simulation-based forecasting experiments and finally, the eighth section concludes.

%SSSSSSSSSSSSSSSSSSSSSSSSSSSSSSSSSSSSSSSSSSSSSSSSSSSSSSSSSSSSSSSSSS
\section{Modeling long memory}

The term strong persistence is used to describe processes, where the more distant past of the process still strongly influences the present, i.e.\ where the autocorrelation function dies out very slowly. Formally, this can be characterized by the autocovariance function $\gamma$ (which is defined as $\gamma(h) = \E \left[ \left(y_t-\mu \right) \left(y_{t+h}-\mu \right) \right]$ for a process $\{y_t\}$ with mean $\mu$) not being summable,

\[
\sum_{h=0}^{H} \gamma(h) \to \  \infty \quad \ \mbox{as } H \to \infty \, ,
\]

\noindent or equivalently by the long run variance $\omega^2 = \sum_{h=-\infty}^{\infty} \gamma(h)$ not being finite.

We will use the terms strong persistence and long memory interchangeably here since a distinction between the two only gets relevant for antipersistent processes, see \cite{Hassler19}.

The most widely used model for strong persistence or long memory is fractional integration of order $d$. We briefly recap the fractionally integrated process $\{y_t\}$ of order $d$, for short $y_t \sim I(d)$, which relies on the fractional integration operator $\Delta^{-d}$ with the usual binomial expansion in terms of the lag operator $L$ and a short memory input process $\{x_t\}$ defined properly below:
\begin{equation} \label{FI}
y_t = \mu +  \Delta^{-d} x_t \, , \quad 0 < d < \frac{1}{2} \, , \ \Delta^{-d} = (1-L)^{-d}= \sum_{j=0}^\infty \pi_j (-d) L^j .
\end{equation} 
A recent discussion of the sequence $\{\pi_j (-d)\}$ is contained in \cite[Sect. 5.3]{Hassler19}. In particular, the coefficients die out as $j^{d-1}/\Gamma(d)$ and are hence not summable for $d>0$. This  causes long memory in that the autocovariances of $\{y_t\}$ at lag $h$  converge to zero at rate $h^{2d-1}$ and hence are not summable, too. This  strength of persistence is sometimes also measured by the rate at which the variance of the cumulated process grows: $\var (\sum_{t=1}^T y_t)$ diverges with $T^{2d +1}$. For these properties to hold true, the input sequence $\{x_t\}$ has to meet certain requirements. We collect them in the following assumption.

\noindent \textbf{Assumption 1} \emph{Let}
 $\{x_t\}$ \emph{be a covariance-stationary  process with}
\[
x_t = c(L) \varepsilon_{t} =  \sum_{j=0}^\infty c_j \varepsilon_{t-j} \, , \quad \E (\varepsilon_t) = 0 \, , \  \E (\varepsilon_t \, \varepsilon_{t+h}) = \left\{ \begin{array}{cc} \sigma^2 \, ,  & h=0 \\ 0 \, , & h \neq 0 \end{array} \right. \, ,
\]
\emph{and with} ($c_0=1$)  %\marginpar{one-summ}
\begin{equation*} \label{one-summ}
\sum_{j=0}^\infty   |c_j| < \infty \quad \mbox{and} \quad c(1) = \sum_{j=0}^\infty c_j \neq 0  \, . %\quad \mbox{and} \quad c_j = o \left( \frac{1}{j}\right)\,  .
\end{equation*}

\noindent The process $\{x_t\}$  is  called integrated of order zero, for short $x_t \sim I(0)$. The restriction   $c(1)  \neq 0 $ implies that the data is not overdifferenced. Consequently, the long-run variance is finite and positive:
\[
0 < \omega_x^2 := \sigma^2 \left(\sum_{j=0}^\infty c_j \right)^2 < \infty \, .
\]
All stationary and invertible ARMA processes meet Assumption 1 and it further holds that $\var (\sum_{t=1}^T x_t)$ grows at  rate $T$.

The inverse filter, the fractional differencing filter,
\[
\Delta^{d} = (1-L)^{d}= \sum_{j=0}^\infty \pi_j (d) L^j \, ,
\]

\noindent removes the persistence or long memory from the FI process: $\Delta^d y_t = x_t$. Note that the mean is removed as well by fractional differencing, $\Delta^d \mu = 0$, because $\sum_{j=0}^\infty \pi_j (d)=0$ holds.

The stationary FI process from (\ref{FI}) is often called of type I since the work by \cite{MarinucciRobinson99}; a truncated version that becomes stationary only asymptotically is called type II. The truncation assumes zero values before $t=1$. We assume such a truncated model instead of (\ref{FI}). The truncated filter becomes
\begin{equation} \label{FI_II}
 \Delta^{-d}_+ x_t := \Delta^{-d} x_t \mathbf{1}_{(t > 0)} (t)  =  \sum_{j=0}^{t-1} \pi_j(-d) {x_{t-j}} \, , \quad t=1,\ldots, T \, ,
\end{equation}
where $\mathbf{1}_{(t > 0)}$ is the usual indicator function:
\[
\mathbf{1}_{(t > 0)} (t) = \left\{ \begin{array}{cc}
1 \, , &  \ t > 0 \\
0 \, , & \ \mbox{else}
\end{array} \right. \, .
\]
Analogously, one defines $\Delta^d_+ x_t := \Delta^d x_t \mathbf{1}_{(t > 0)} (t)$, such that by construction $\Delta^d_+ \Delta^{-d}_+ =  \mathbf{1}_{(t > 0)} $ or $\Delta^d_+ \Delta^{-d}_+ x_t = x_t  \mathbf{1}_{(t > 0)} (t)  $. Truncated fractional differencing and integration has been discussed more formally e.g.\ by \cite{Johansen08}.  Note that a truncated filter also allows to define truely nonstationary FI processes, i.e. that for the process
\begin{equation} \label{FI_typeII}
y_t = \mu +  \Delta_+^{-d} x_t \, ,  
\end{equation}
$d$ is unrestricted, in particular $d \geq \frac{1}{2}$ is not ruled out. This is important as we do not want to rule out nonstationary processes when it comes to forecasting.

Being equipped with a model that can be employed when forecasting under strong persistence, the issues of how to estimate the long memory parameter $d$ and the mean $\mu$ come up naturally. These will be discussed in the next two sections.

\section{Estimating the memory parameter}

Due to the inherent difficulties in estimating the long memory parameter $d$ there is a large literature on this topic. A natural first choice for an estimator is of course the exact maximum likelihood [ML] estimator (see e.g.\ \cite{Dahlhaus89} and \cite{Dahlhaus06}), which unfortunately has several deficiencies in this context, including its restriction to the stationary region, the required gaussianity assumption and the need for estimating the mean. %and difficulties in the computation of the covariance matrix. 
The first two problems can be overcome by the conditional sum of squares estimator (see e.g. \cite{Beran95}, \cite{HualdeRobinson11} or \cite{Nielsen15}), which is a time domain approximation to exact ML, while the Whittle estimator (see e.g.\ \cite{FoxTaqqu86} or \cite{VelascoRobinson00}) additionally solves the third problem, being a frequency domain approximation, which makes estimation of the mean unnecessary. Nevertheless, all of these estimators require the full specification of the whole model. 

As assuming knowledge of the full model might be quite heroic, semiparametric estimators have been proposed as an alternative to these fully parametric estimators. The most popular class of semiparametric estimators are local or narrowband estimators, which only use frequency domain information, i.e.\ the periodogram, in a vicinity of the origin, where the long memory component of the process dominates. They consequently only estimate the long memory parameter $d$ and not all the parameters of a fully specified model. The Geweke Porter Hudak [GPH] estimator (see e.g.\ \cite{GPH83} and \cite{HDB98}) and the local Whittle [LW] estimator (see e.g.\ \cite{Kunsch87} and \cite{RobinsonLW}) and some of its variants as the most prominent representatives of this class are settled in the frequency domain and thus do not require estimation of the mean, too. While the GPH estimator follows a simple idea and is very easy to implement, the LW estimator does not require a normality assumption and is more efficient. For these estimators, a bandwidth $m$ has to be chosen, balancing a high variance caused by staying too close to the origin and using too little information and a bias induced by the contamination of the estimation through the short memory component of the process. As data-driven bandwidth selection does not work well in practice (see e.g.\ \cite{HurvichDeo99}, \cite{Henry01} or \cite{AndrewsGuggenberger03}), ad hoc bandwidth choices as a function of the sample size $T$ are often used. We will also follow this road. For example, the bandwidth for the LW estimator is usually chosen as $m = O(T^{\alpha})$ with $0 < \alpha < 1$. Besides the choice of the bandwidth, the price to pay for the local semiparametric estimators is a slower rate of decay of the asymptotic variance of $O(\frac 1 m)$ compared to $O(\frac 1 T)$ for the parametric estimators. For example, the approximate variance of the local Whittle estimator is $\frac{1}{4m}$. Regarding finite sample performance of the estimators, \cite{NielsenFrederiksen05} performed a systematic comparison within the classes of parametric and local semiparametric estimators through simulations. From the parametric estimators they recommend the use of the Whittle estimator as it clearly dominates its time domain competitors. For the semiparametric estimators, they find that the local Whittle estimator and some of its variants as for example the local polynomial Whittle (see \cite{AndrewsSun04}) outperform the GPH estimator and its variants.

\begin{figure}
	\centering
	\includegraphics[width=0.6\linewidth]{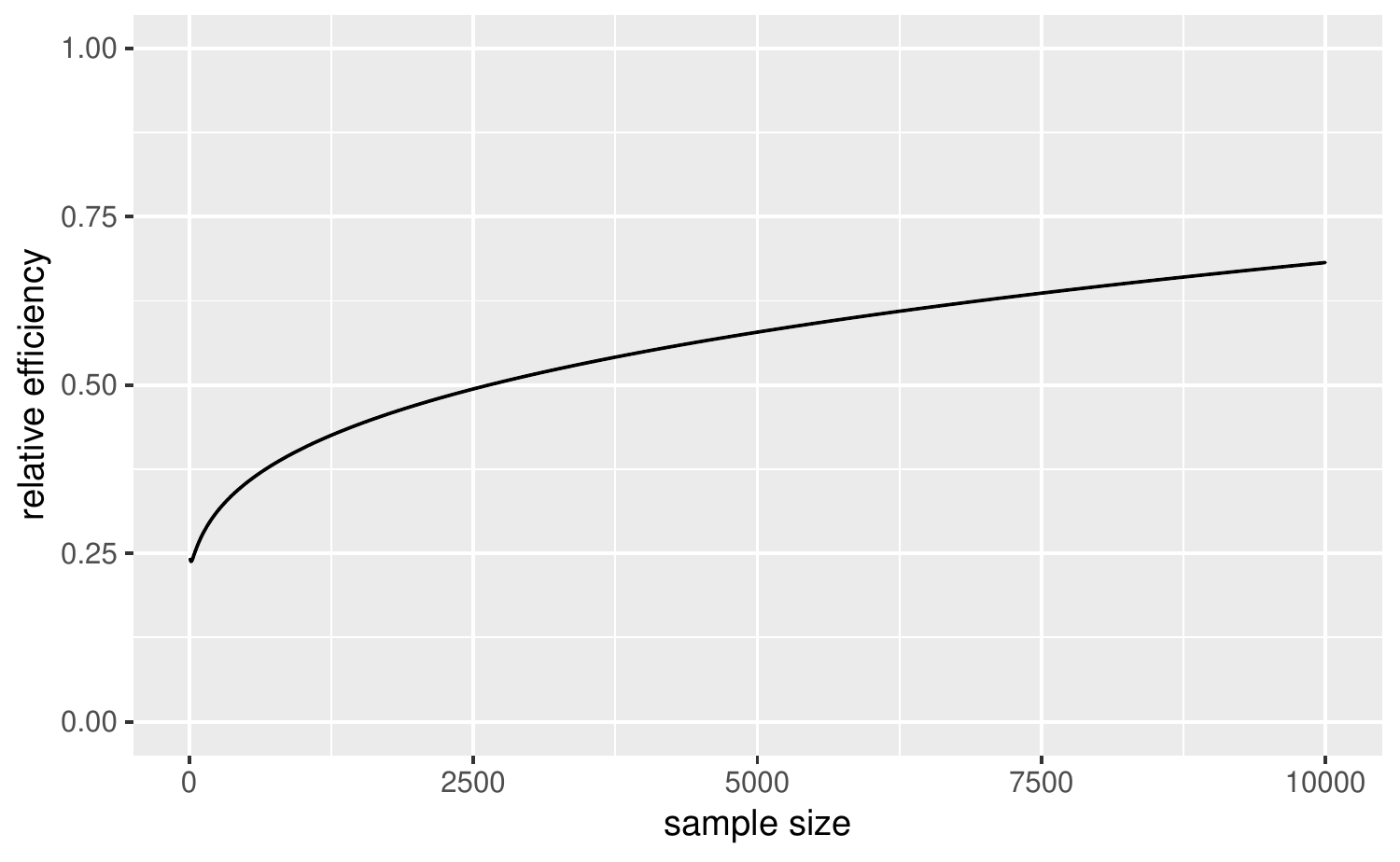}
	\caption[]{asymptotic efficiency of $\hat{d}_{LW}$ with bandwidth $m= \lfloor T^{0.65}\rfloor $ relative to that of $\hat{d}_{GSW}$ with lag length $k=\lfloor \log(T) \rfloor$}
	\label{fig:refflwvsgw}
\end{figure}

So-called global or broadband semiparametric estimators represent an alternative to the other two classes. As their name suggests, they use information over the whole frequency range, but their semiparametric nature comes in through the approximation of the short memory component via a certain parametric model of an order which is growing with the sample size. The global semiparametric Whittle [GSW] estimator proposed by \cite{BhansaliGiraitisKokoszka06} uses an AR($p$) model here, while an alternative global estimator works with a log-periodogram regression of a fractional exponential model (see \cite{MoulinesSoulier99}, \cite{MoulinesSoulier00} and \cite{HurvichBrodsky01}). These estimators have the nice property that their asymptotic variance decays with a faster rate than the one of the semiparametric estimators. For example, the GSW has an approximate variance of $\frac p T$, where $p = O(\log(T))$ has been recommended. Nevertheless, a closer comparison of the approximate variances of the two classes of estimators reveals a different picture in finite samples. Figure~\ref{fig:refflwvsgw} shows the efficiency of the LW relative to the GSW estimator, where the widely used choice $m= \lfloor T^{0.65} \rfloor$ for the bandwidth and the recommended choice $p=\lfloor \log(T) \rfloor$ for the lag length are picked. Thus, for all reasonable sample sizes, the uncertainty associated with the local Whittle estimator is considerably smaller. To the best of our knowledge, the finite sample performance of global semiparametric estimators has not been systematically analyzed yet. For more details on the estimators and their properties see e.g.\ \cite{Beranetal13} or \cite{Hassler19} and further references therein.

When it comes to forecasting specifically, a wide array of different estimators have been used in literature. There seems to be no consensus on which estimator to use, on the contrary, there is an ongoing debate. \cite{Baillieetal12} advocate the use of the maximum likelihood estimator, which they compare to the local Whittle. In a comment to this \cite{arteche2012comment} stresses that their results are unsurprising as they compared the performance of a parametric and a semiparametric estimator under a known model and did not really touch upon the issues of model misspecification and model selection. 

Arising from the above considerations are three basic types of feasible estimators: Locally semiparametric estimators, globally semiparametric estimators and parametric estimators plus a model selection step. Unfortunately, we are not aware of any work that compares these three types of estimation procedures or two of them. To do this, the model selection step for the parametric estimators or at least an analysis of the consequences of model misspecification would have to be included. Furthermore, we are not aware of any analysis of the finite sample performance of global semiparametric estimators. To clarify these open issues before choosing estimators for forecasting, we perform a small simulation study. As touched upon above, the Whittle and the local Whittle estimator are favorable in terms of theoretical properties and finite sample performance within the classes of parametric and local semiparametric estimators respectively. Consequently, we use these two as representatives of the respective classes.\footnote{For the class of semiparametric estimators, we could have used variants of the local Whittle as well, e.g.\ the local polynomial Whittle, the fully extended local Whittle (see \cite{Abadiretal07}) or the exact local Whittle estimator (see \cite{ShimotsuPhillips05} and \cite{Shimotsu10}), which can even lead to improvements in some circumstances. In a recent study, \cite{CheungHassler2018} present experimental evidence from a systematic comparison of six variants of Whittle type estimators. Depending on the range of $d$ they all have their merits and drawbacks, but overall there is no estimator dominating local Whittle uniformly. As it would not have been clear which from these estimators to choose for general forecasting purposes and as a large scale comparison of estimators is not within the scope of this paper, we use the basic variant of the local Whittle.} As the global semiparametric Whittle estimator is theoretically more appealing than the log-periodogram regression of the fractional exponential model, i.e.\ being more efficient and not needing a normality assumption, and as there are no studies concerning their finite sample performance, we use the GSW as a representative of the global semiparametric estimators in our study.
 
Even though consistency and limiting normality for the GSW estimator have only been established for the stationary region and under a restricted optimization window using a pre-estimator, we do not use this restriction and a pre-estimator since this would exclude the nonstationary region, which is not practical. The remaining procedure is very simple: An ARFI($p$,$d$) model is used, where $p=\lfloor \log(T) \rfloor$ and then the Whittle objective function is minimized to estimate $d$. The procedure for the Whittle estimator is very similar: We again assume an ARFI($p$,$d$) model, pick $p$ by Akaike's information criterion\footnote{We set the maximum lag order to $p_{max}=\lfloor \log(T) \rfloor$, too.} (see \cite{BeranBhansaliOcker98}) and then estimate $d$ by minimizing the respective Whittle objective function. We use the AIC instead of alternative selection criteria as it is the usual choice in a forecasting context. For the local Whittle estimator, we pick the bandwidth as $m = \lfloor T^\alpha \rfloor$ with $\alpha \in \{0.5, 0.65, 0.8\}$, which is standard in the literature.

As data generating processes we use FI($d$) processes of type II, i.e.\ generated from model (\ref{FI_typeII}), with two different values of $d$, $d \in \{0.4,0.7\}$, and three different input processes for $\{x_t\}$ from Assumption 1 with increasing persistence. We measure their degree of persistence by the variance ratio of long-run variance to variance,  $VR=\omega_x^2/\gamma_x(0)$. For standard normal iid innovations $\{\varepsilon_t\}$, the models are
\[
iid: \ x_t = \varepsilon_t \quad \mbox{with } VR=1 \, ,
\]
\[
AR(1): \ x_t = 0.5 \, x_{t-1} + \varepsilon_t \quad \mbox{with } VR=3 \, ,
\]
\[
MA(9): \ x_t = \varepsilon_t + \frac{9}{10} \varepsilon_{t-1} + \cdots + \frac{1}{10} \varepsilon_{t-9} \quad \mbox{with } VR=7.857 \, .
\]

In Table \ref{d_estimation_0.4} we present the mean squared errors (MSEs) for $d=0.4$ using 1000 repetitions from estimating $d$ with the local Whittle estimator with bandwidth $m = \lfloor T^\alpha \rfloor$ [$LW(T^\alpha)$], the global semiparametric Whittle estimator and the Whittle estimator plus AIC [W(AIC)]. For comparison we also present the Whittle estimator for ARFI($p$, $d$) models for different values of $p$ fixed a priori [W($p=p^*$)]. We report results for four different sample sizes $T \in \{60,300,1500,7500\}$. The smallest MSE in each row is marked in bold face.

\begin{table}[h]
	\centering
	\caption{MSEs of different estimators for $d$ for simulated FI(0.4) processes with short memory component $x_t$ and sample size $T$, 1000 repetitions}
	\label{d_estimation_0.4}
	\begin{adjustbox}{width=\textwidth}
		\begin{tabular}{@{}llllllllllllllll@{}}
			\toprule
			$x_t$ & $T$  & $LW (T^{0.5})$ & $LW (T^{0.65})$ & $LW (T^{0.8})$ & $GSW$ & $W (AIC)$   & $W (p=0)$ & $W (p=1)$   & $W(p=3)$ & $W (p=10)$ \\ \midrule
			iid & 60  &0.1155&0.0405&\textbf{0.0171}&10.9716&11.4711&0.0212&0.4328&4.3762&14.0576\\
			& 300 &0.0291&0.0092&0.0036&0.9612&0.7556&\textbf{0.0026}&0.02&0.2101&2.1613\\
			& 1500 &0.0094&0.0027&0.0009&0.0331&0.0226&\textbf{0.0005}&0.0013&0.0051&0.0815\\
			& 7500 &0.0037&0.0009&0.0002&0.0015&0.0005&\textbf{0.0001}&0.0002&0.0005&0.002\\
			\\
			AR(1) & 60 &0.1322&\textbf{0.1135}&0.151&9.863&10.0318&0.1752&0.1357&4.0548&11.2652\\
			& 300 &0.0316&\textbf{0.0265}&0.0966&1.6752&1.6751&0.1746&0.0532&0.7963&2.2098\\
			& 1500 &0.0099&\textbf{0.005}&0.05&0.0674&0.0534&0.1725&0.0074&0.0167&0.1243\\
			& 7500 &0.0035&0.0011&0.0229&0.0018&0.0011&0.1721&\textbf{0.0007}&0.0009&0.0022\\
			\\
			MA(9) & 60 &0.5093&0.6676&0.5636&8.4441&7.4402&0.6814&\textbf{0.0722}&3.0257&10.1518\\
			& 300 &0.0438&0.3271&0.62&1.8444&1.4754&0.7735&\textbf{0.0176}&0.8141&1.5582\\
			& 1500 &0.0101&0.0393&0.5601&0.061&0.0665&0.8047&\textbf{0.0097}&0.1352&0.1845\\
			& 7500 &\textbf{0.0038}&0.0041&0.3875&0.0058&0.005&0.8144&0.0101&0.0107&0.0186\\
			\bottomrule
		\end{tabular}
	\end{adjustbox}
\end{table}

As expected, for the iid short memory component the correctly specified Whittle estimator with $p=0$ performs best. The local Whittle is clearly the best of the feasible estimators, irrespective of the bandwidth choice. Nevertheless, a larger bandwidth is of course a better choice in this case as there are no short memory dynamics, which could cause a bias. Very surprising at the first glance are the extremely bad results of the Whittle estimator plus AIC and the global semiparametric Whittle estimator. Only for $T=7500$ their MSEs have the same order of magnitude as the ones of the local Whittle estimators. For the more persistent short memory components the same picture emerges regarding the comparison of the three feasible estimators. For the local Whittle estimator the smaller bandwidth choices lead to improvements due to the increased short memory dynamics. A look at the last four columns of the table helps to understand why the Whittle estimator plus AIC and the global semiparametric Whittle estimator perform so badly. This seems to be caused by the drastic drop in performance of the Whittle estimator under even slight misspecification. For example if the model order is picked as 1 instead of 0 or 0 instead of 1, the accuracy most often drops below the one of the local Whittle estimator. For larger magnitudes of misspecification, the accuracy continues to decrease strongly.\footnote{An exception here is that the Whittle estimator with $p=1$ performs quite well for the process with MA(9) innovations. This is probably due to the similarity of this MA(9) process to an AR(1) process with a coefficient of about 0.9.} Thus, as the AIC or the deterministic order choice by the global semiparametric Whittle estimator often do not pick exactly the correct model, this slight misspecification in combination with the bad performance of the Whittle estimator under slight misspecification lead to these results.

\begin{table}[h]
	\centering
	\caption{{MSEs of different estimators for $d$ for simulated FI(0.7) processes with short memory component $x_t$ and sample size $T$, 1000 repetitions}}
	\label{d_estimation_0.7}
	\begin{adjustbox}{width=\textwidth}
		\begin{tabular}{@{}llllllllllllllll@{}}
			\toprule
			$x_t$ & $T$  & $LW (T^{0.5})$ & $LW (T^{0.65})$ & $LW (T^{0.8})$ & $GSW$ & $W (AIC)$   & $W (p=0)$ & $W (p=1)$   & $W(p=3)$ & $W (p=10)$ \\ \midrule 
			iid & 60 &0.11185&0.03803&\textbf{0.01858}&11.39186&11.56437&0.01929&0.41093&4.37725&14.32783\\
			& 300 &0.02869&0.00969&0.00389&0.94002&0.63425&\textbf{0.0027}&0.0176&0.20995&2.11467\\
			& 1500 &0.01002&0.00317&0.00106&0.02898&0.01844&\textbf{0.0006}&0.00143&0.00431&0.07296\\
			&7500 &0.00421&0.0012&0.00028&0.00168&0.0008&\textbf{0.00014}&0.00031&0.00071&0.00211\\
			\\
			AR(1) & 60 &0.12669&\textbf{0.09768}&0.10569&10.63713&9.72827&0.14552&0.16326&4.16247&12.31798\\
			& 300 &0.03075&\textbf{0.02805}&0.08284&1.51804&1.23276&0.15889&0.05856&0.71517&2.14612\\
			& 1500 &0.01097&\textbf{0.00658}&0.04769&0.06314&0.03568&0.16459&0.0077&0.01453&0.11269\\
			& 7500 &0.0043&0.00171&0.02288&0.00201&0.00148&0.1689&\textbf{0.00103}&0.00107&0.00247\\
			\\
			MA(9) & 60 &0.3995&0.42793&0.30085&11.03323&8.03117&0.40488&\textbf{0.13162}&3.8611&13.02868\\
			& 300 &\textbf{0.04445}&0.25644&0.40295&2.19134&1.18821&0.5054&0.08068&0.68322&1.93014\\
			& 1500 &\textbf{0.01161}&0.04029&0.45088&0.07743&0.08606&0.61862&0.0383&0.10376&0.16573\\
			& 7500 &\textbf{0.00438}&0.00494&0.35516&0.00528&0.00568&0.70029&0.01046&0.00698&0.01254\\
			\bottomrule
		\end{tabular}
	\end{adjustbox}
\end{table}

For different strengths of long memory the conclusions do not change as can be seen from Table \ref{d_estimation_0.7}, which contains the results for nonstationary FI processes with $d=0.7$.

The results of our simulation study suggest that global semiparametric estimators and parametric estimators with the prior use of a model selection criterion should be used with care. Only for very large sample sizes a reasonable performance can be expected. It is a safer choice to stick with a local semiparametric estimator like the local Whittle, which we will do throughout the rest of the paper.

%SSSSSSSSSSSSSSSSSSSSSSSSSSSSSSSSSSSSSSSSSSSSSSSSSSSSSSSSSSSSSSSSS
\section{Estimating the mean}

Estimating the mean is naturally quite difficult under strong persistence since long trending spells of the time series away from its mean are common. This is reflected in the behavior of the sample mean $\overline y = T^{-1} \sum_{t=1}^T y_t$. From the behavior of $\var (\sum_t y_t)$ given above, it is clear that  $\overline y$ converges all the more slowly to $\mu$ the longer the memory is, see \cite{SamarovTaqqu88}. To see if improvements over $\overline y$ are possible, in this section we analyze the behavior of the alternative mean estimators proposed by \cite{Robinson94EffTests} and \cite{Shimotsu10} theoretically as well as through simulations.

Note that an application of the truncated fractional difference filter to (\ref{FI_typeII}) results in a time-varying mean:
\begin{equation} \label{LM_inv}
\Delta_+^{d} y_t = r_t \mu  + x_t \, , \quad r_t= \sum_{j=0}^{t-1} \pi_j(d) \, .
\end{equation}
This suggests to follow the proposal by \cite{Robinson94EffTests} and to regress the filtered data $\Delta_+^{d} y_t$ on the variable $r_t$ to determine $\widehat \mu$ by ordinary least squares [OLS] from (\ref{LM_inv}). Such an estimation is justified by the maximum likelihood principle if $\{x_t\}$ is normal white noise. \cite{Robinson94EffTests} establishes that removing $r_t \widehat \mu$  does not affect the limiting distribution of his subsequent test. It seems that a comparison of this estimator with the arithmetic mean has never been carried out.

First, we determine the limiting variance under stationarity (more precisely: under $d< 1/2$) in order to evaluate the efficiency.

\begin{proposition} \label{prop_Asym_Var}
	Assume  Assumption 1 in model (\ref{LM_inv}) for $0 < d < 1/2$. It then holds that
	\[
	 {T^{1-2d}} \var(\widehat \mu) \ \to \ \frac{\omega_x^2 (1-2d)(\pi d)^2}{\Gamma^2 (d+1) \sin^2 (\pi d)}
	\]
	as $T \to \infty$, where $\Gamma$ is the gamma function.
\end{proposition}

\noindent{\sc Proof} See Appendix.

\begin{figure}
	\centering
	\includegraphics[width=0.6\linewidth]{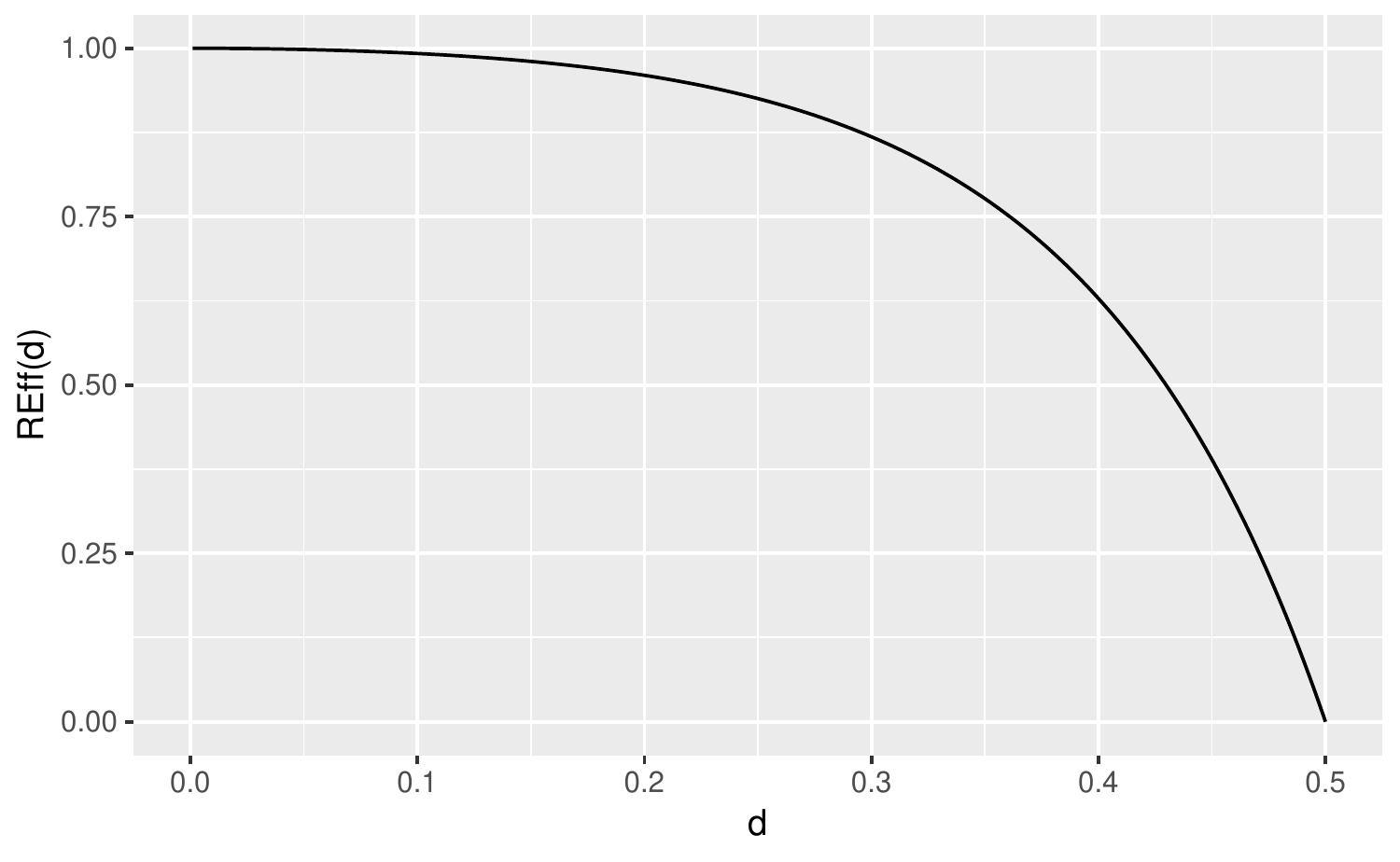}
	\caption{$REff(d)$ according to Proposition \ref{prop_Asym_Var}}
	\label{fig_REff}
\end{figure}

The same rate applies for the arithmetic mean. Under mildly stronger assumptions \cite[Coro.\ 2.3]{Tanaka99} found that
\[
T^{1-2d} \var (\overline y) \ \to \ \frac{\omega_x^2}{\Gamma^2 (1+d) (2d+1)} \, ,
\]
see also \cite[Thm.\ 1]{MarinucciRobinson00}. Let us define as relative efficiency
\[
	REff (d) := \lim_{T \to \infty} \frac{\var (\widehat \mu)}{\var (\overline y)} = (1-4 d^2 )\frac{(\pi d)^2}{\sin^2 (\pi d)}\, .
\]
The graph of this function in Figure \ref{fig_REff} shows that there is plenty of room for efficiency gains over the arithmetic mean for larger values of $d$, at least asymptotically. Note that this result, which is obtained under the assumption of a type II process, which is often argued to be more realistic and better suited for empirical analysis, see e.g.\ \cite{JohansenNielsen16}, is very different from the result for type I processes obtained by \cite{Adenstedt74}, where hardly any efficiency gains over the arithmetic mean are possible, see \cite{SamarovTaqqu88}. Furthermore, note that $\widehat \mu$ is a function of $d$, $\widehat \mu = \widehat \mu (d)$, since the filter $\Delta_+^d$ depends on $d$. A preliminary estimation $\widehat d$ is required to construct a feasible estimator $\widehat \mu (\widehat d)$.

Next, we turn to the nonstationary case. It is clear that the sample mean  diverges for $d \geq 1/2$: In fact, we know from \citet[Thm.\ 1]{MarinucciRobinson00} that $\var (\overline y)$ diverges with $T^{2d-1}$ for $d> 1/2$. Note that the variance of $\widehat \mu$ from Proposition 1 also vanishes only under $d < 1/2$.  For nonstationary fractionally integrated processes the estimator is, however, still defined, and its variance is
\begin{eqnarray*}
\var(\widehat \mu (d)) = \frac{\gamma_x (0) \sum_{t=1}^T r_t^2 + 2 \sum_{h=1}^{T-1} \gamma_x (h) \sum_{t=1}^{T-h} r_t r_{t+h}}{\left( \sum_{t=1}^T r_t^2 \right)^{2}} .
\end{eqnarray*}
For $d > 1/2$ the regressors $r_t$ converge to zero so quickly that $\sum_{t=1}^T r_t^2$ remains finite for growing $T$ and $\mu$ is not estimated consistently, but, contrary to $\overline y$, the estimator at least has a finite variance.

\citet{Shimotsu10} considered the starting value as an estimator under nonstationarity, $\widetilde \mu = y_1$ with $\widetilde \mu - \mu = x_1$. Note that $\widehat \mu (1) = \widetilde \mu$. For general $d$, the ratio of the variances of these two estimators is 
\[
\frac{\var(\widehat \mu (d))}{\var(\widetilde \mu)} = \frac{1}{\sum_{t=1}^T r_t^2} + 2 \, \sum_{h=1}^{T-1} \frac{\gamma_x (h)}{\gamma_x (0)}\frac{ \sum_{t=1}^{T-h} r_t r_{t+h}}{\left( \sum_{t=1}^T r_t^2 \right)^2}\, .
\]
If $\{x_t\}$ was white noise, then ${\var(\widehat \mu (d))}/{\var(\widetilde \mu)} = {1}/{\sum_{t=1}^T r_t^2} \leq 1$, but under serial correlation of $\{x_t\}$ it is not clear which estimator is more precise. The full proposal of Shimotsu is to use the arithmetic mean, the starting value or a linear combination of both, depending on the value of $d$:

\[
\widetilde{\mu} (d)  =  v(d) \, \overline y + (1-v(d)) \, y_1 \, ,
\]
where
\[
v(d) = \left\{ \begin{array}{cc} \vspace{0.2 cm} 1 \, , & d \leq 1/2 \\ \vspace{0.2 cm} \frac{1+ \cos(4 \pi  d)}{2} \, , & 1/2 <  d < 3/4 \\  0 \, , & d
\geq 3/4. \end{array} \right. 
\]

A feasible version of this estimator $\widetilde{\mu} (\widehat d)$ also requires a preliminary estimation of the long memory parameter.

To assess how the theoretical findings concerning these three estimators carry over to practical performance and how the feasible estimators work, we conducted a simulation study. For the estimation of $d$ we used the local Whittle estimator with bandwidth $m=\lfloor T^{0.65} \rfloor$. As data generating processes, we used again FI($d$) processes, i.e.\ model (\ref{FI_typeII}), with the same three short memory input processes $\{x_t\}$ with increasing persistence as before. We used three different values of $d$, namely $d \in \{0.2,0.4,0.6\}$, and as sample sizes we used again $T \in \{60, 300, 1500, 7500\}$. Table \ref{mu_estimation} presents the MSEs over 10000 repetitions, where at each instance the estimator with the smallest MSE is marked in bold face again. Overall, the estimator proposed by Robinson dominates the other two, while Shimotsu's estimator usually shows the second best performance and the arithmetic mean the worst. The gains over the arithmetic mean can be quite large and are rising with $d$ and with the persistence of the short memory component. Only for $d=0.2$ and the smaller sample sizes the arithmetic mean outperforms the other two estimators. But in general, it seems to be worth to accept the increased uncertainty coming with the estimation of $d$ in the first step of the feasible estimators to overall get a more precise estimate of the mean. 

\begin{table}[]
	\centering
	\caption{MSEs of different estimators for $\mu$ for simulated FI($d$) processes with short memory component $x_t$ and sample size $T$, $d$ is estimated by LW with $m=\lfloor T^{0.65} \rfloor$, 10000 repetitions}
	\label{mu_estimation}
	\begin{adjustbox}{width=\textwidth}
		\begin{tabular}{@{}llllllllllllll@{}}
			\toprule
			\multirow{2}{*}{d} & \multirow{2}{*}{$x_t$} & \multicolumn{3}{l}{$T=60$} & \multicolumn{3}{l}{$T=300$} & \multicolumn{3}{l}{$T=1500$} & \multicolumn{3}{l}{$T=7500$} \\ \cmidrule(r){3-5} \cmidrule(r){6-8} \cmidrule(r){9-11} \cmidrule(r){12-14} \\
			&               &    $\overline y$     & Shimotsu          &Robinson  &    $\overline y$      & Shimotsu          &Robinson &   $\overline y$     & Shimotsu          &Robinson &    $\overline y$      & Shimotsu          &Robinson        \\ \midrule
			$0.2$ & iid  &\textbf{0.0714}&0.0792 &0.0843 & \textbf{0.0281}& \textbf{0.0281}  &0.0285  &0.0107&0.0107  &\textbf{0.0104}  &0.0041&0.0041  &\textbf{0.0039}   \\
			&      AR(1)  &\textbf{0.291}&0.4799  &0.4733    &\textbf{0.1073}&0.108  &0.1275   &0.0417&0.0417  &\textbf{0.0408}  &0.0158&0.0158  &\textbf{0.0153}    \\
			&       MA(9)  &\textbf{2.1529}&3.765  &3.6557   &\textbf{0.8372} & 3.3225  &3.1316 &\textbf{0.3194}&0.3195  &0.4502 &0.1217&0.1217  &\textbf{0.1196}     \\
			$0.4$ & iid &0.307&0.3383 &\textbf{0.3043}  &0.229&0.2283  &\textbf{0.1904} &0.1655&0.1654 &\textbf{0.1232} &0.1195&0.1195  &\textbf{0.0827}    \\
			&      AR(1)  &1.248&1.1351  &\textbf{1.0047} &0.8761&0.7844  &\textbf{0.6641}  &0.6445&0.6364  &\textbf{0.442} &0.4669&0.4669  &\textbf{0.3155}     \\
			&       MA(9)  &9.2067&3.8866  &\textbf{3.6284}  &6.8461&3.9072  &\textbf{3.9044}  &4.9629&\textbf{3.2068}  &3.2374 &3.602&3.597  &\textbf{2.2469}    \\
			$0.6$ & iid &1.2811&0.9747  &\textbf{0.7306} &1.8127&1.054 &\textbf{0.6361} &2.4894&1.1748 &\textbf{0.5631} &3.4138&1.4086  &\textbf{0.5388}     \\
			&      AR(1)  &5.1844&1.5449  &\textbf{1.4219}  &6.925&1.5802  &\textbf{1.374}  &9.6651&2.2303  &\textbf{1.4011}  &13.3503&3.9409  &\textbf{1.459}   \\
			&       MA(9) &38.0374&3.9581  &\textbf{3.3317}  &54.1704&3.9209  &\textbf{3.514}  &74.725&\textbf{3.9517}  &5.1493  &103.2586&12.5138  &\textbf{7.1468}     \\
			\bottomrule
		\end{tabular}
	\end{adjustbox}
\end{table}

Hence, we will work with $\widehat \mu (d)$ for a given $d$ and with $\widehat \mu (\widehat d)$ for estimated $\widehat d$ in the next sections when it comes to forecasting real and simulated series using FI models since this estimator is consistent for $d< 1/2$, has a finite variance for larger orders of integration and shows a good performance in our simulations.

An alternative strategy would be to ignore the mean and simply work with $ \Delta_+^d y_t$   because $\Delta_+^d 1 \to \Delta^d 1 =0$ for $t \to \infty$: $ \Delta_+^d y_t  \approx  x_t$. 
Clearly, this cannot be recommended since $\Delta_+^d 1 =   \sum_{j=0}^{t-1} \pi_j(d)$ for small $d$ converges very slowly with $t$ getting large.

%SSSSSSSSSSSSSSSSSSSSSSSSSSSSSSSSSSSSSSSSSSSSSSSSSSSSSSSSSSSSSSSSSSSSS
\section{Forecasting methods}

In this section we present in detail how we use the FI model to construct forecasting methods for strongly dependent time series and we describe the relevant competing methods, which do not employ long memory models. We assume a stretch of data of size $T$, $y_1, \ldots, y_T$, used to forecast $h$ steps ahead: $\widehat{y}_{T+h}$.

We use methods that explicitly account for long memory in that they (try to) remove the strong persistence from the series by filtering with a fractional difference filter, forecast the filtered data, and then recolor these forecasts by applying the fractional integration filter. Because of the findings from the previous section, the mean is accounted for from the filtered data by Robinson's method. With model (\ref{LM_inv}) in mind we hence proceed as follows for a given $d$. Compute $ \Delta^d_+ y_t$ in order to estimate $\widehat \mu$, and save the residuals:
\[
\xi_t := \Delta^d_+ y_t - r_t \widehat \mu \, , \quad t=1, \ldots, T \, .
\]
They are fed into an autoregression  on $p$ lags to account for short memory in $\{x_t\}$,  where $p$ is determined by Akaike's information criterion, AIC:\footnote{Note that no intercept is included as the original series has already been demeaned. Furthermore, the inclusion of an intercept clearly deteriorated the forecasting performance in our experiments. The maximum lag length considered is $12 \lfloor (T/100)^{0.25}\rfloor$.}   
\[
\xi_t =  \widehat{a}_1 \xi_{t-1} + \ldots + \widehat{a}_p \xi_{t-p} + \widehat{\varepsilon}_t \, , \quad t=p+1,\ldots, T.
\]
The autoregressive scheme is then used to recursively forecast $h$ steps ahead:\footnote{Consider some forecasted sequence $\widehat{z}_{T+h-j}$. Whenever $T+h-j \leq T$, then $\widehat{z}_{T+h-j} = {z}_{T+h-j}$.}
\begin{equation} \label{filter_forecast}
\widehat{\xi}_{T+h} := \widehat{a}_1 \widehat{\xi}_{T+h-1} + \ldots  + \widehat{a}_p \widehat{\xi}_{T+h-p} \, .
\end{equation}
Finally,  forecasts of the original sequence are given by
%\footnote{Because of $V_+ (L) F_+(L) =  \mathbf{1}_{(t > 0)} $ this is equivalent to computing the recursion
%\[
%\widehat{y}_{T+h} = - \sum_{j=1}^{T+h-1} v_j \widehat{y}_{T+h-j} + \widehat{\xi}^{T+h} \, .
%\]}
\begin{equation} \label{y_hat_LM}
\widehat{y}_{T+h} = \widehat \mu +  \Delta_+^{-d} \widehat{\xi}_{T+h} \, .
\end{equation}
Because of $\Delta_+^{-d} \Delta_+^{d} =  \mathbf{1}_{(t > 0)} $, this two-step procedure is equivalent to computing the recursion
\[
\widehat{y}_{T+h} = - \sum_{j=1}^{T+h-1} \pi_j(d) \widehat{y}_{T+h-j} + r_t \widehat \mu + \widehat{\xi}_{T+h} \, ,
\]
which is also often applied when forecasting with FI models.

Before the fractional differencing filter $\Delta_+^{d}$ can be applied, a value of $d$ has to be chosen. The estimation of the long memory parameter has been discussed in the third section. Resulting from our analysis of the different estimators there, we use the local Whittle estimator with bandwidth $m$ that we pick as $m = \lfloor T^\alpha \rfloor$ with  $\alpha \in \{0.5, 0.65, 0.8\}$ (In the tables throughout the next two chapters this method is abbreviated by FI($T^{\alpha})$). We use three different bandwidth choices to analyze if the forecasting performance strongly depends on the bandwidth.

Even though the local Whittle estimator yields good estimation results compared to other estimators, it still has a high variance due to the inherent difficulty of estimating the long memory parameter. Motivated by this, we propose the unorthodox method of using a fixed long memory component in the forecasting method, i.e.\ a FI model with an a priori fixed parameter $d$, to circumvent the estimation problem. A natural choice here is to set $d=0.5$ as this is the border to nonstationarity and somehow the middle between the classical $I(0)$ and $I(1)$ paradigms. Furthermore, the dynamics of many economic time series are believed to lie somewhere close to the borderline between stationarity and nonstationarity. The exact choice of $d$ does probably not matter too much anyway - as long as it is not as far away from the truth as the outcomes of classical long memory estimators may often be - since the flexible short memory component in form of the AR model may adapt to the fixed long memory component to a certain degree. Thus, using a FI($0.5$) model within the forecasting method (which will also be abbreviated by FI($0.5$)) may be a good middle ground, robustifying the forecasting procedure against strong persistence by being able to account for slowly decaying autocorrelations, but not running into problems with the very volatile long memory estimators. 

One classical competitor of the true fractionally integrated model is its special case from the $I(1)$ world, at which we arrive by setting $d=1$. The FI($1$) model, which is an ARI($p$,$1$) model when using the method described above, plays an important role in modeling and forecasting economic time series since the seminal work by \cite{NelsonPlosser82} and is thus a relevant competitor, appearing in the long memory forecasting literature since \cite{SmithYadav94}. %Furthermore, integer differencing robustifies against breaks in mean.
Another FI($1$) model that is a classical benchmark and a sanity check is the no change forecast, $\widehat{y}_{T+h} = y_T$, arising from a random walk, which we also included in our experiments. As it was dominated by the other procedures, we do not report its results in the tables throughout the rest of the paper.

The fiercest competitor from the literature is certainly the autoregressive model. Since \citet{Ray1993}, it has been repeatedly argued that modeling fractional integration does not provide better forecasts than  simply fitting an autoregression, in this context often called  long  autoregression. The autoregression  (now with intercept) of the unfiltered series (as we are in the special case of a FI($0$) model) builds again on a lag length $p$ determined by  AIC:
\[
y_t =  \widehat{\nu} + \widehat{a}_1 y_{t-1} + \ldots + \widehat{a}_p y_{t-p} + \widehat{\varepsilon}_t \, , \quad t=p+1,\ldots, T \, .
\]

Autoregressions are the classical benchmark in forecasting economic time series, being simple and at the same time very hard to beat, see e.g.\ \cite{StockWatson1998comparison}. For forecasting under long memory this competitor is even more relevant due to the repeated empirical and simulation-based claims that forecasts based on LAR models would be superior to forecasts based on FI models and even theoretical claims that they would be able to capture true long memory. For a clarification and a recent theoretical underpinning of the latter see \citet{DemetrescuHassler2016}. As a simplification of the LAR model without the need of selecting a lag length, an often used benchmark is the AR(1) model, which we also included in our experiments. As it is dominated by the other procedures, we refrain from reporting its results, too.

Besides the autoregressive forecasts, the second dominant benchmark method used in the general forecasting literature is exponential smoothing. It is known to perform well as a highly flexible general-purpose forecasting method for a wide variety of situations, see e.g.\ \cite{DeGooijer}. Some theoretical work has provided a solid foundations  for the method in form of state space models (see \cite{hyndman2002state}) as well as demonstrated that exponential smoothing arises as the optimal method in certain highly relevant situations, for example under certain forms of structural change, see e.g.\ \cite{ClementsHendry2006} and \cite{pesaran2013optimal}. Thus, we also include exponential smoothing in our forecasting experiments even though it has traditionally not been used in the long memory forecasting literature.\footnote{We use the ets function from the R forecast package, where a trend component is chosen or not (additive or multiplicative, with or without dampening) according to (corrected) AIC. We do not use a seasonal component. The reference for the forecast package is: Hyndman R., G. Athanasopoulos, C. Bergmeir, G. Caceres, L. Chhay, M. O'Hara-Wild, F. Petropoulos, S. Razbash, E. Wang, F. Yasmeen (2019). forecast: Forecasting functions for time series and linear models. R package version 8.5, http://pkg.robjhyndman.com/forecast. }
	
Finally, we also used the arithmetic mean, $\widehat{y}_{T+h} = \overline y$, as a standard benchmark and sanity check and to get an idea up to which forecasting horizon the forecasts from the other methods contain valuable information.

%SSSSSSSSSSSSSSSSSSSSSSSSSSSSSSSSSSSSSSSSSSSSSSSSSSSSSSSSSSSSSSSSSSSS
\section{Empirical evidence}

In this section we present the results of systematic pseudo-out-of-sample [POOS] forecasting experiments on real time series with the methods described in the previous section to assess their predictive performance. We use inflation and realized volatility series as these are classical examples of real-world long memory time series from economics and finance. We present robustness checks in several directions, changing amongst others the time period and the country. Finally, we also assess the effect of the estimation window size and compare rolling to expanding window forecasts. 

\subsection{Inflation}

Inflation is the classical example of a macroeconomic variable exhibiting long memory. Since \cite{HaWo95} and \cite{BaillieChungTieslau96} many papers have analyzed inflation time series using long memory models. Surprisingly, in the long memory forecasting literature, there is not much work on inflation forecasting.

\begin{figure}
	\centering
	\begin{subfigure}[b]{0.45\textwidth}
		\centering
		\includegraphics[width=1\textwidth]{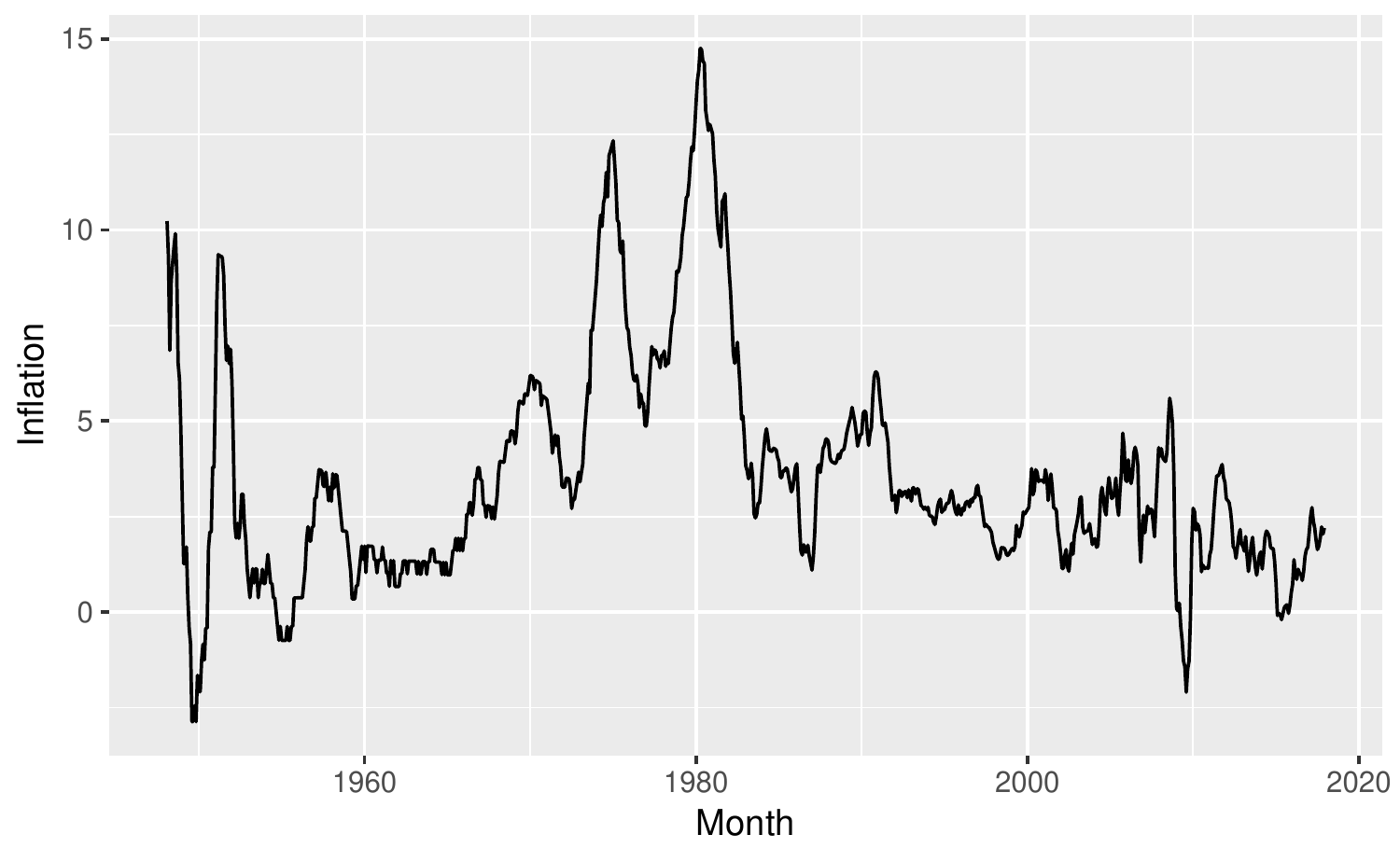}
		\caption{timeplot}
		\label{fig:timeplotyoyinflation}
	\end{subfigure}
	\begin{subfigure}[b]{0.45\textwidth}
		\centering
		\includegraphics[width=1\textwidth]{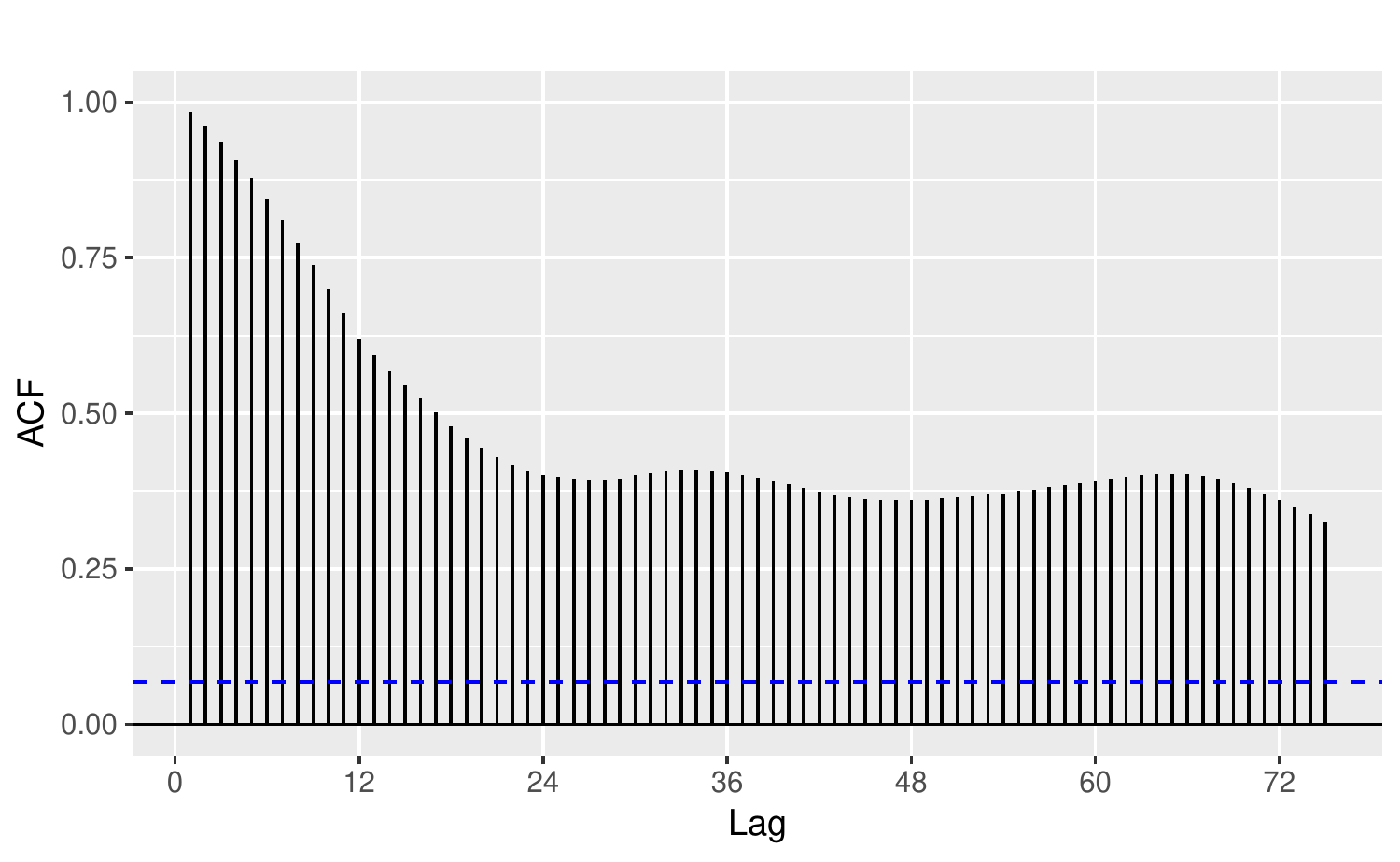}
		\caption{autocorrelogram}
		\label{fig:correlogramyoyinflation}
	\end{subfigure}
	\caption{plots for monthly year-on-year US CPI inflation from January 1948 to November 2017}
	\label{fig:plotyoyinflation}
\end{figure}

Firstly, we use monthly year-on-year US consumer price inflation.\footnote{We calculated the inflation from the US consumer price index without seasonal adjustment. The exact reference for the series is: U.S. Bureau of Labor Statistics, Consumer Price Index for All Urban Consumers: All Items [CPIAUCNS], retrieved from FRED, Federal Reserve Bank of St. Louis; https://fred.stlouisfed.org/series/CPIAUCNS, January 3, 2018.}  The series spans from January 1948 to November 2017 and is shown in part (a) of Figure \ref{fig:plotyoyinflation}. The autocorrelogram of the series shown in part (b) of this figure decays very slowly as is characteristic for long memory time series. We ran a pseudo-out-of-sample forecasting experiment with a rolling estimation window of size 180 for six different forecasting horizons $h \in \{1,3,6,12,24,48\}$. The predictive performance of the competing methods is measured by the mean squared forecast error. In Table \ref{infl1}, we report relative mean squared forecast errors in relation to the method based on the FI($0.5$) model. The smallest relative mean squared error is marked in bold face, the second smallest in italics. 

Regarding the results, a first striking observation is that the FI($0.5$) method uniformly outperforms all methods based on short memory models over all horizons. From these short memory methods only LAR shows a reasonable performance with losses between 2 and 17 percent compared to FI($0.5$), which are increasing with the forecasting horizon. This suggests that modeling long memory explicitly when forecasting under long memory seems to be beneficial even in this very simple form of a fixed long memory component. Furthermore, when forecasting further into the future, this seems to become more and more important. 

A second striking and at first rather surprising observation concerns the performance of the methods based on a FI model with estimated $d$: Even though we have taken great care in choosing a well-performing estimator compared to the existing alternatives with the local Whittle, the FI($0.5$) method again almost uniformly outperforms FI($T^{0.5}$), FI($T^{0.65}$) and FI($T^{0.8}$). Thus, irrespective of the choice of the bandwidth, here the fixed-$d$ method is clearly superior to the methods with estimated long memory component and the gap widens with increasing forecasting horizon. The latter observation will be analyzed in detail in the next section with the help of simulations. As discussed in the previous section, including a fixed long memory part in a forecasting method could thus provide a good way to robustify against long memory, while not being plagued by the problems usually arising when one tries to estimate the exact strength of the memory. Although we have chosen $d=0.5$, picking any value between 0 and 1 for $d$ is of course possible and can lead to improvements upon the limiting cases LAR and FI($1$), which both perform already quite well.

Thirdly, looking further at the performance of the estimated-$d$ long memory methods, they perform better than the best short memory method, i.e.\ LAR, irrespective of the bandwidth chosen. The only exception is the very large forecast horizon $h=48$, where their performance strongly drops off. Comparing the three different bandwidth choices no pattern emerges which would indicate an optimal bandwidth choice here, even though the performance clearly varies with the bandwidth. This indicates that choosing a bandwidth here is indeed a difficult task.

\begin{table}[h]
	\centering
	\caption{MSEs relative to FI($0.5$) from POOS forecasting experiments for monthly year-on-year US CPI inflation, January 1948 to November 2017, rolling window of size 180}
	\label{infl1}
	\begin{adjustbox}{width=0.75\textwidth}
		\begin{tabular}{@{}llllllll@{}}
			\toprule
			horizon & FI ($T^{0.5}$) &  FI ($T^{0.65}$) &  FI ($T^{0.8}$)   & FI ($1$)   & LAR  & ES & Mean  \\ \midrule
			$h=1$ &\textit{1.01}&1.015&\textbf{1.007}&1.026&1.021&1.426&74.922\\
			$h=3$ &\textbf{1.007}&\textit{1.019}&1.021&1.033&1.037&1.252&13.686\\
			$h=6$ &\textbf{0.998}&\textit{1.008}&1.022&1.022&1.065&1.167&5.602\\
			$h=12$ &1.038&\textit{1.022}&1.036&\textbf{1.015}&1.078&1.152&2.268\\
			$h=24$ &1.157&\textit{1.063}&\textit{1.063}&\textbf{1.012}&1.169&1.34&1.3\\
			$h=48$ &1.802&1.306&1.206&\textbf{1.11}&\textit{1.16}&1.819&1.292\\
			\bottomrule
		\end{tabular}
	\end{adjustbox}
\end{table}

To find out whether these observations continue to hold more generally, we first take a look at several robustness checks, then move on to another type of time series, namely financial time series in the form of realized volatility, and finally use simulated series, too. We executed robustness checks in several directions, starting with a shorter time span of the original inflation series, then switching from year-on-year to month-on-month inflation and finally changing the country, looking at German inflation. Furthermore, at the end of this section we look at an analysis of the effects of the size of the rolling window and of the use of an expanding window. Table \ref{infl2} includes the results for the shortened series starting from 1970 to analyze the effects of the specific time period used. Qualitatively nothing changes and all observations from above continue to hold.

\begin{table}[h]
	\centering
	\caption{MSEs relative to FI($0.5$) from POOS forecasting experiments for monthly year-on-year US CPI inflation, January 1970 to November 2017, rolling window of size 180}
	\label{infl2}
	\begin{adjustbox}{width=0.8\textwidth}
		\begin{tabular}{@{}llllllll@{}}
			\toprule
			horizon & FI ($T^{0.5}$) &  FI ($T^{0.65}$) &  FI ($T^{0.8}$)   & FI ($1$)   & LAR & ES & Mean   \\ \midrule
			$h=1$ &\textbf{1.01}&1.024&\textit{1.011}&1.035&1.024&1.518&41.128\\
			$h=3$ &\textbf{1.004}&\textit{1.035}&1.037&1.054&1.038&1.314&7.008\\
			$h=6$ &\textbf{0.986}&\textit{1.021}&1.044&1.048&1.074&1.195&3.037\\
			$h=12$ &\textbf{1.065}&1.1&1.127&\textit{1.097}&1.115&1.252&1.635\\
			$h=24$ &1.282&\textit{1.203}&1.248&\textbf{1.115}&1.379&1.337&1.596\\
			$h=48$ &1.513&1.299&1.231&\textbf{0.962}&1.085&\textit{1.034}&1.194\\
			\bottomrule
		\end{tabular}
	\end{adjustbox}
\end{table}

\begin{figure}
	\centering
	\begin{subfigure}{0.45\textwidth}
		\centering
		\includegraphics[width=1\linewidth]{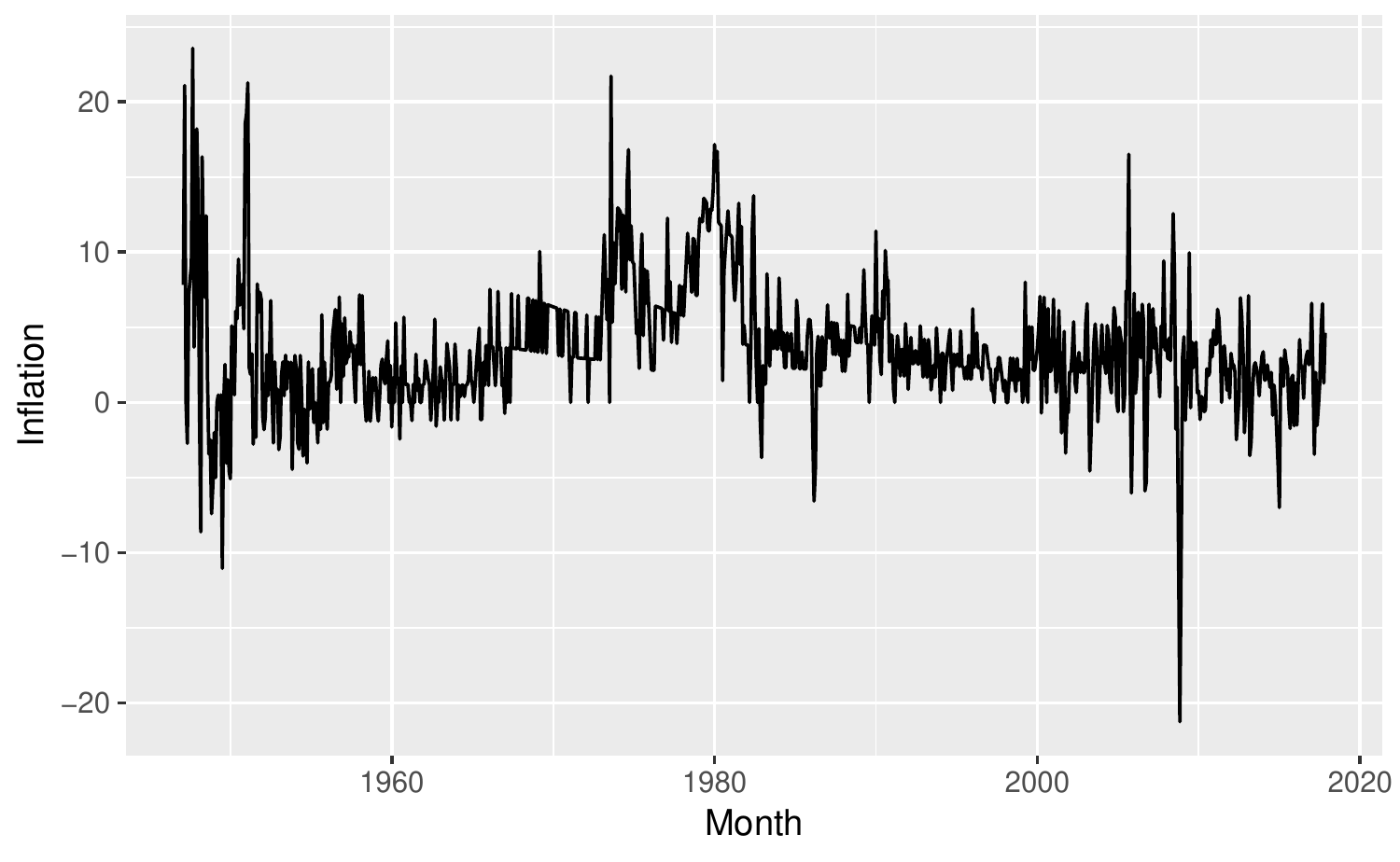}
		\caption{timeplot}
		\label{fig:timeplotmominflation}
	\end{subfigure}
	\begin{subfigure}{0.45\textwidth}
		\centering
		\includegraphics[width=1\linewidth]{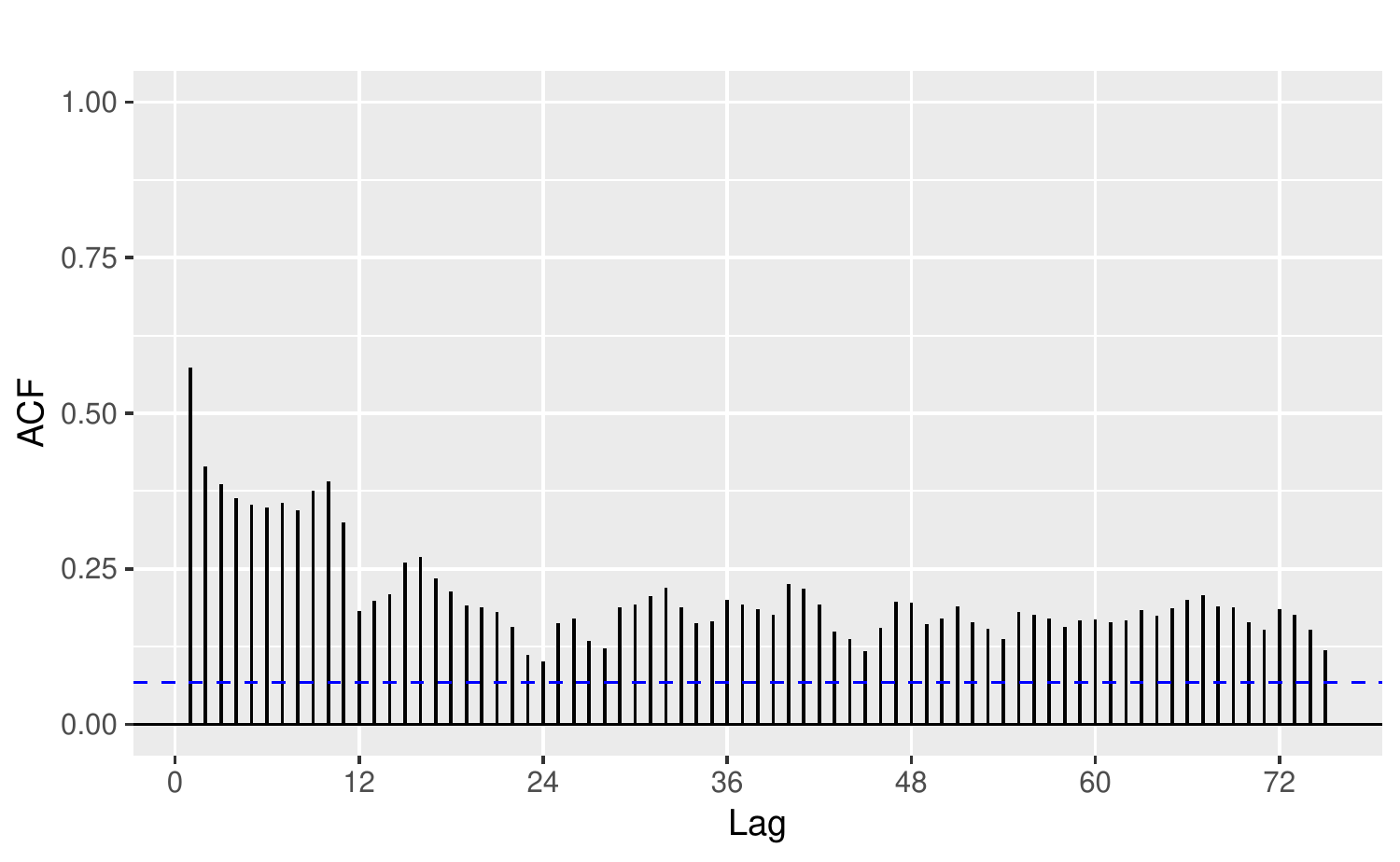}
		\caption{autocorrelogram}
		\label{fig:correlogrammominflation}
	\end{subfigure}
	\caption{plots for annualized month-on-month US CPI inflation from January 1948 to November 2017}
	\label{fig:plotmominflation}
\end{figure}

In the next step we change the definition of inflation and look at annualized month-on-month inflation from January 1948 to November 2017.\footnote{We calculated the inflation from the seasonally adjusted US consumer price index. The exact reference for the series is: U.S. Bureau of Labor Statistics, Consumer Price Index for All Urban Consumers: All Items [CPIAUCSL], retrieved from FRED, Federal Reserve Bank of St. Louis; https://fred.stlouisfed.org/series/CPIAUCSL, January 3, 2018.} Naturally, month-on-month inflation is much more volatile and less persistent than year-on-year inflation, which is visible in Figure \ref{fig:plotmominflation}. Despite the quite different characteristics of the series, the pseudo-out-of-sample forecasting results presented in Table \ref{infl3} are very similar: FI(0.5) is still clearly the best method, even though the methods with estimated $d$ now perform better than before. Furthermore, exponential smoothing now performs much better than before even though still not better than the long autoregression or the long memory methods.

\begin{table}[h]
	\centering
	\caption{MSEs relative to FI($0.5$) from POOS forecasting experiments for annualized month-on-month US CPI inflation, February 1947 to November 2017, rolling window of size 180}
	\label{infl3}
	\begin{adjustbox}{width=0.8\textwidth}
		\begin{tabular}{@{}llllllll@{}}
			\toprule
			horizon & FI ($T^{0.5}$) &  FI ($T^{0.65}$) &  FI ($T^{0.8}$)   & FI ($1$)   & LAR & ES & Mean  \\ \midrule
			$h=1$ &1.018&\textbf{1.003}&\textbf{1.003}&1.033&1.013&1.044&1.652\\
			$h=3$ &1.027&\textit{0.999}&1.013&1.084&1.022&\textbf{0.988}&1.348\\
			$h=6$ &1.024&\textbf{1}&\textit{1.008}&1.074&1.048&1.064&1.399\\
			$h=12$ &\textit{1.02}&\textbf{1.008}&1.023&1.022&1.05&1.073&1.216\\
			$h=24$ &\textit{1.008}&\textbf{0.993}&1.028&1.109&1.147&1.175&1.094\\
			$h=48$ &\textbf{1.001}&\textit{1.003}&1.02&1.042&1.256&1.088&1.14\\
			\bottomrule
		\end{tabular}
	\end{adjustbox}
\end{table}

\begin{figure}
	\centering
	\begin{subfigure}{0.45\textwidth}
		\centering
		\includegraphics[width=1\linewidth]{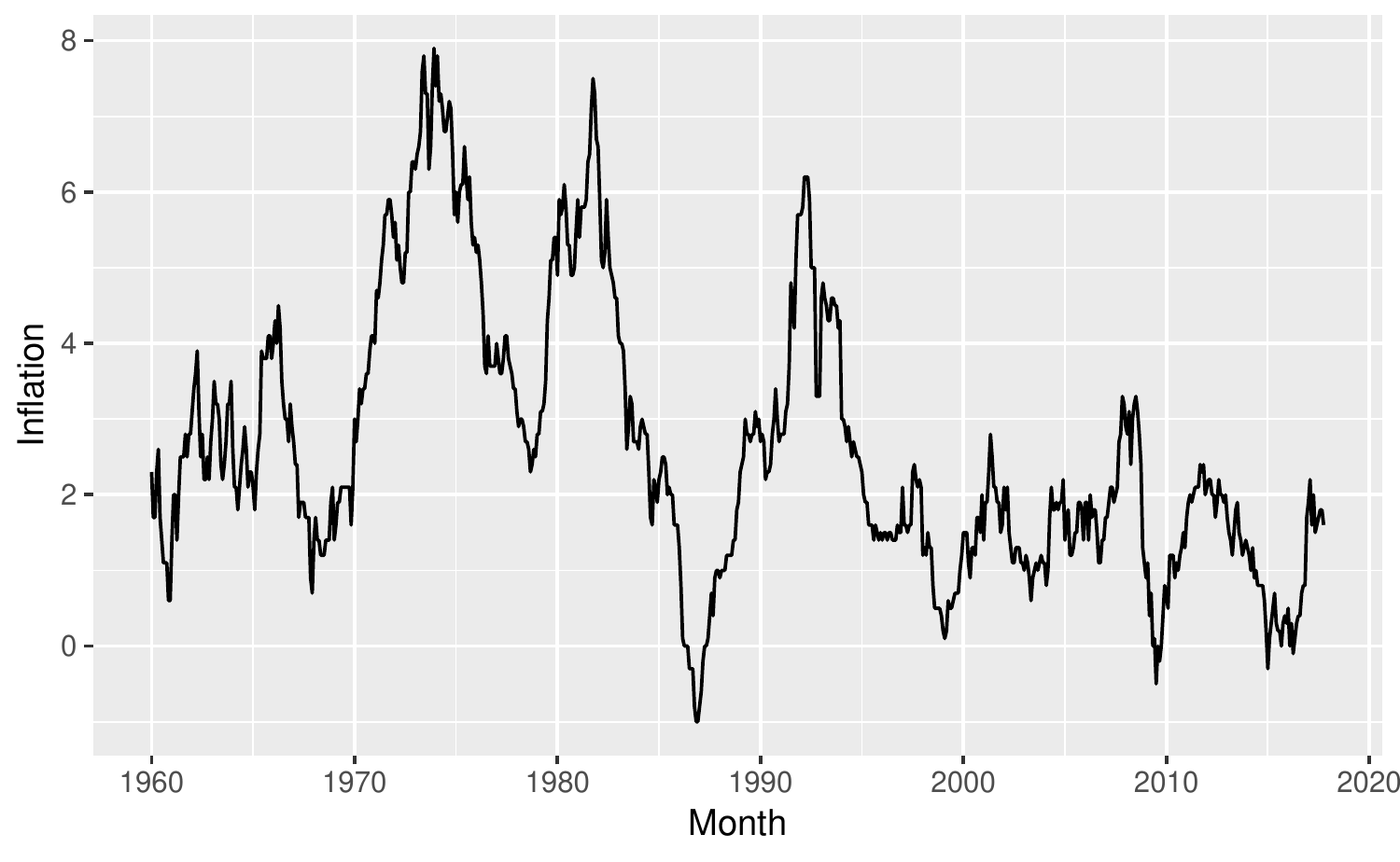}
		\caption{timeplot}
		\label{fig:timeplotyoyinflation_Germany}
	\end{subfigure}
	\begin{subfigure}{0.45\textwidth}
		\centering
		\includegraphics[width=1\linewidth]{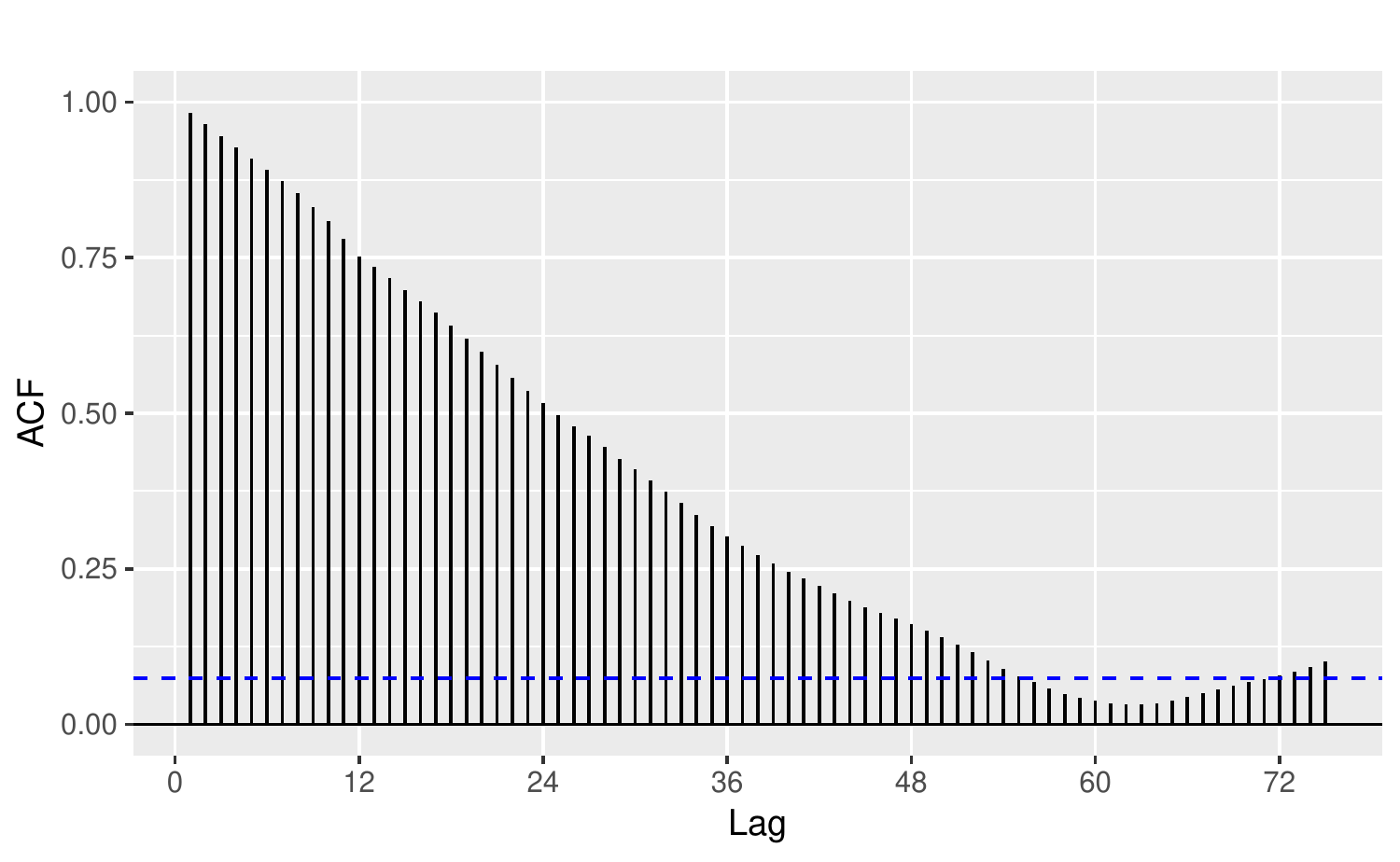}
		\caption{autocorrelogram}
		\label{fig:correlogramyoyinflation_Germany}
	\end{subfigure}
	\caption{plots for monthly year-on-year German CPI inflation from January 1960 to November 2017}
	\label{fig:plotyoyinflation_Germany}
\end{figure}

To check robustness with respect to the country, we use German year-on-year inflation,\footnote{Here, the exact reference is: Organization for Economic Cooperation and Development, Consumer Price Index: OECD Groups: All Items for Germany [CPALTT01DEM659N], retrieved from FRED, Federal Reserve Bank of St. Louis; https://fred.stlouisfed.org/series/CPALTT01DEM659N, February 8, 2018.} which is depicted in Figure \ref{fig:plotyoyinflation_Germany}. The results reported in Table \ref{infl4} are similar again: FI(0.5) outperforms all other methods by a margin increasing in horizon and often being very large.\footnote{This does of course not hold true for the arithmetic mean, which in relative terms gets better as the informational content of the conditioning information used by the other methods gets smaller and smaller.} What is different here is that the LAR method performs better than the FI($T^{\alpha}$) methods. Such a behavior could explain the mixed evidence reported in the literature concerning the usefulness of the FI model for forecasting: Due to the high uncertainty in estimating $d$, here the LAR method seems to be superior to the long memory methods if we only look at the ones with estimated $d$. But by using the fixed-$d$ method, we can resolve this issue as we clearly see that modeling the long memory explicitly without having to estimate its strength leads to improved forecasting results.

\begin{table}[h]
	\centering
	\caption{MSEs relative to FI($0.5$) from POOS forecasting experiments for monthly year-on-year German CPI inflation, January 1960 to November 2017, rolling window of size 180}
	\label{infl4}
	\begin{adjustbox}{width=0.8\textwidth}
		\begin{tabular}{@{}llllllll@{}}
			\toprule
			horizon & FI ($T^{0.5}$) &  FI ($T^{0.65}$) &  FI ($T^{0.8}$)   & FI ($1$)   & LAR  & ES & Mean \\ \midrule
			$h=1$ &1.01&\textbf{1.003}&1.011&1.014&\textit{1.005}&1.161&28.975\\
			$h=3$ &1.041&1.032&\textit{1.028}&1.039&\textbf{1.012}&1.103&8.808\\
			$h=6$ &1.076&1.061&\textit{1.059}&1.069&\textbf{1.033}&1.093&4.158\\
			$h=12$ &1.168&1.148&1.126&1.137&\textit{1.07}&\textbf{1.065}&1.796\\
			$h=24$ &1.364&1.283&\textit{1.256}&1.262&\textbf{1.214}&1.297&1.272\\
			$h=48$ &1.849&1.588&1.527&1.495&\textit{1.343}&1.522&\textbf{0.979}\\
			\bottomrule
		\end{tabular}
	\end{adjustbox}
\end{table}

\subsection{Realized Volatility}

Another area where many time series show long memory characteristics and the FI model is widely used is finance. A prime example of long memory time series in this field are return volatility series. Since the rise of high frequency data in finance and the introduction of realized volatility measures (see e.g.\ \cite{AndersenBollerslev98} or \cite{Andersenetal01}), the latter are widely used in financial volatility forecasting, being more accurate measures of volatility and better proxies of unobserved conditional variances than squared returns (see e.g.\ \cite{HansenLunde2011}). FI models are natural candidates for forecasting realized volatility and since \cite{Andersenetal03} have often been applied for that purpose and compared to the performance of competing methods (see again \cite{HansenLunde2011} for a literature overview). A model that is also widely employed in forecasting realized volatility is the HAR model by \cite{corsi2009}. The idea behind the model is that different volatility components arise over different time horizons (trading days, weeks and months) due to different types of market participants and their trading behavior. Being just a restricted AR model, it is very simple and shows a good forecasting performance and has thus become a workhorse model for volatility forecasting. Consequently, we add it as a very important competitor to the forecasting experiments for realized volatility. 

\begin{figure}
	\centering
	\begin{subfigure}{0.45\textwidth}
		\centering
		\includegraphics[width=1\linewidth]{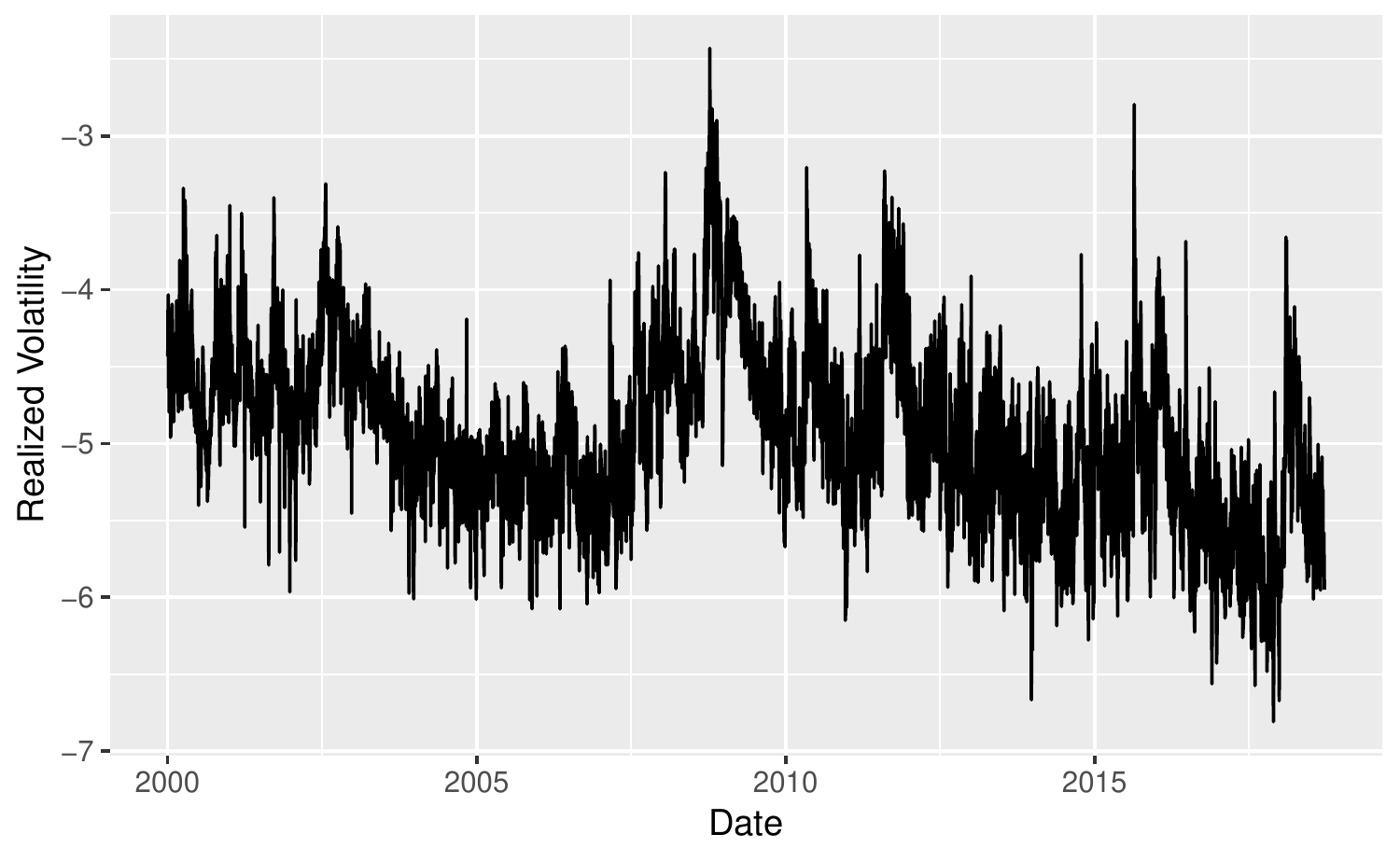}
		\caption{timeplot}
		\label{fig:timeplotRV}
	\end{subfigure}
	\begin{subfigure}{0.45\textwidth}
		\centering
		\includegraphics[width=1\linewidth]{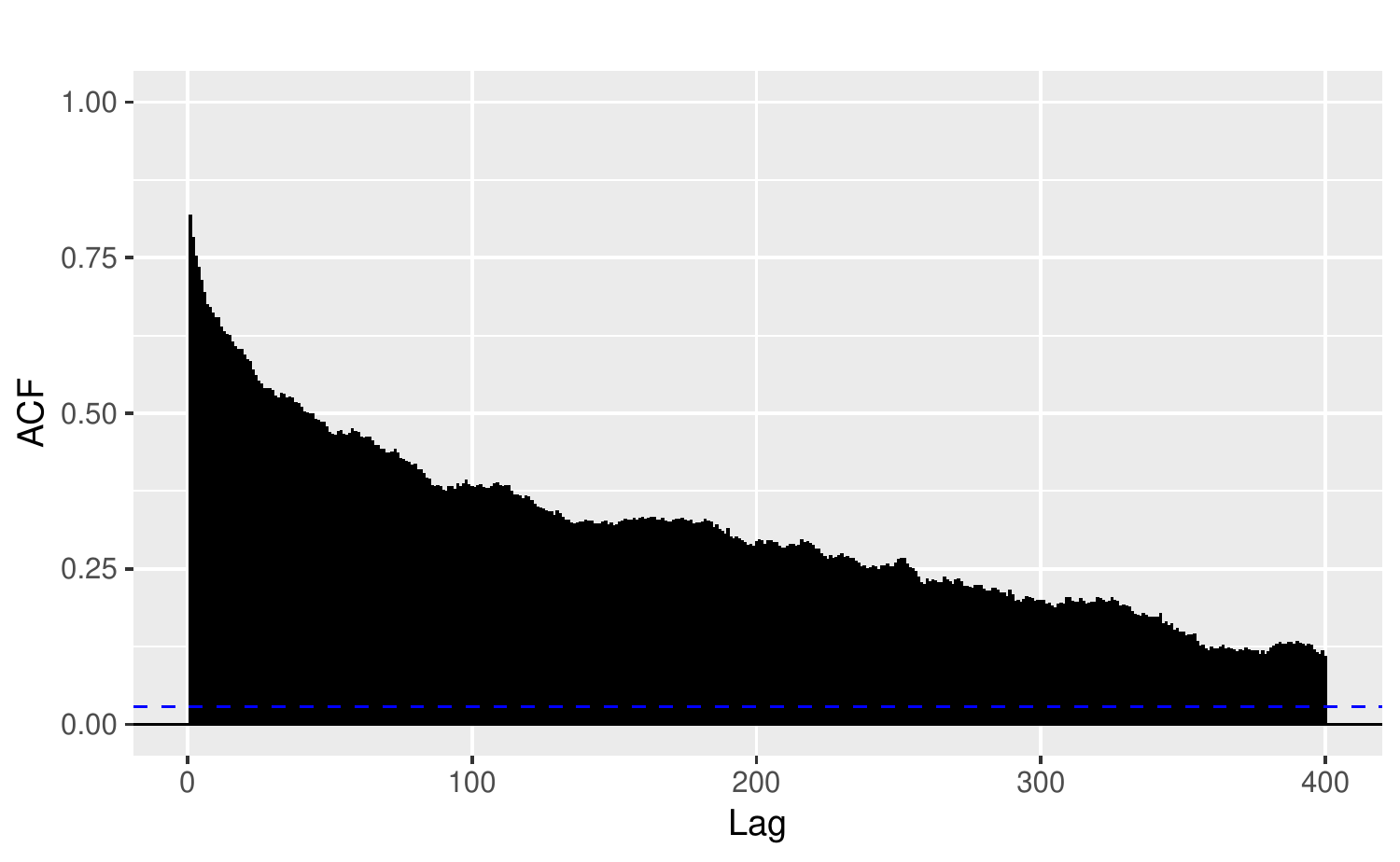}
		\caption{autocorrelogram}
		\label{fig:correlogramRV}
	\end{subfigure}
	\caption{plots for daily S\&P 500 realized volatility $\log(\widehat{\sigma}_t)$ from 3.01.2000 to 20.09.2018}
	\label{fig:plotRV}
\end{figure}

A standard way to measure realized volatility is to use the five-minute sub-sampled realized variance, $\widehat{\sigma}_t^2$,  or its square root, $\widehat{\sigma}_t$, respectively (for details see e.g.\ \cite{shephard2010}). To reduce skewness and kurtosis, usually logs of this measure are taken. We follow these conventions and use the logs of the square root of the five-minute sub-sampled realized variance, $\log(\widehat{\sigma}_t)$, of the S\&P 500 from January 2000 to September 2018 and execute a pseudo-out-of-sample forecasting experiment with a rolling window of size 500 on it.\footnote{This time series as well as the other realized variance series stem from the ``Oxford-Man Institute's realized library, version 0.3'', Oxford-Man Institute, University of Oxford by Heber G., A. Lunde, N. Shephard and K. Sheppard (2009), retrieved September 25, 2018.} The series and its autocorrelogram are depicted in Figure \ref{fig:plotRV}. The extremely slowly decaying autocorrelation function clearly indicates long memory. Although, apart from that, its characteristics are quite different from the inflation series, the results shown in Table \ref{RV1} are very similar: The short memory methods are uniformly outperformed by FI(0.5) and the gap tends to get bigger with the forecasting horizon. Now, the forecasts based on the HAR model are the best forecasts based on a short memory model, showing a decent performance, especially for short horizons. With regards to the long memory methods, here the procedures with estimated $d$ show slight improvements of up to two percent over FI(0.5). This difference compared to the inflation experiments may be due to the now less complicated dynamics of the underlying short memory process, which makes the estimation of $d$ easier. More light will be shed on this question through the simulations in the next section.

\begin{table}[h]
	\centering
	\caption{MSEs relative to FI($0.5$) from POOS forecasting experiments for daily realized volatility $\log(\widehat{\sigma}_t)$ of the S\&P 500, from 3.01.2000 to 20.09.2018, rolling window of size 500}
	\label{RV1}
	\begin{adjustbox}{width=0.8\textwidth}
		\begin{tabular}{@{}lllllllll@{}}
			\toprule
			horizon  & FI ($T^{0.5}$) &  FI ($T^{0.65}$) &  FI ($T^{0.8}$)  & FI ($1$)   & LAR  & ES & Mean & HAR \\ \midrule
			$h=1$ &1.001&\textit{0.998}&\textbf{0.995}&1.026&1.017&1.018&3.043&1.004\\
			$h=3$ &1&\textbf{0.997}&\textbf{0.997}&1.046&1.038&1.056&2.32&1.011\\
			$h=5$ &0.998&\textbf{0.995}&\textit{0.997}&1.063&1.054&1.096&2.028&1.009\\
			$h=10$ &1.002&\textbf{0.994}&\textit{0.998}&1.078&1.087&1.13&1.756&1.019\\
			$h=20$ &0.992&\textbf{0.984}&\textit{0.991}&1.106&1.118&1.147&1.505&1.019\\
			$h=40$ &\textit{0.982}&\textbf{0.98}&0.993&1.136&1.124&1.182&1.334&1.032\\
			$h=80$ &\textbf{0.98}&\textit{0.981}&0.987&1.181&1.114&1.227&1.209&1.055\\
			\bottomrule
		\end{tabular}
	\end{adjustbox}
\end{table}

To check robustness with respect to the time period and especially to the financial crisis, we also executed the analysis on the latter part of the series starting in 2012. As the results in Table \ref{RV2} show, no large changes occur. Now, there are virtually no differences between the FI ($T^{\alpha}$) methods and FI($0.5$) and HAR performs a little worse than before.

\begin{table}[h]
	\centering
	\caption{MSEs relative to FI($0.5$) from POOS forecasting experiments for daily realized volatility $\log(\widehat{\sigma}_t)$ of the S\&P 500, from 2.01.2012 to 20.09.2018, rolling window of size 500}
	\label{RV2}
	\begin{adjustbox}{width=0.8\textwidth}
		\begin{tabular}{@{}llllllllll@{}}
			\toprule
			horizon & FI ($T^{0.5}$) &  FI ($T^{0.65}$) &  FI ($T^{0.8}$)  & FI ($1$)   & LAR & ES & Mean & HAR \\ \midrule
			$h=1$ &\textit{1.003}&1.005&\textbf{1}&1.026&1.019&1.053&2.884&1.015\\
			$h=3$ &\textit{1.001}&1.004&\textbf{0.999}&1.061&1.043&1.13&1.924&1.031\\
			$h=5$ &\textbf{1}&1.004&\textit{1.002}&1.087&1.064&1.215&1.618&1.022\\
			$h=10$ &1.004&\textbf{0.999}&\textbf{0.999}&1.109&1.102&1.277&1.439&1.032\\
			$h=20$ &\textit{0.994}&0.995&\textbf{0.99}&1.177&1.118&1.297&1.286&1.032\\
			$h=40$ &1.003&\textit{1.002}&\textbf{0.999}&1.23&1.105&1.377&1.198&1.049\\
			$h=80$ &1.037&\textit{1.015}&\textbf{0.991}&1.391&1.107&1.545&1.118&1.133\\
			\bottomrule
		\end{tabular}
	\end{adjustbox}
\end{table}

\begin{figure}
	\centering
	\begin{subfigure}{0.45\textwidth}
		\centering
		\includegraphics[width=1\linewidth]{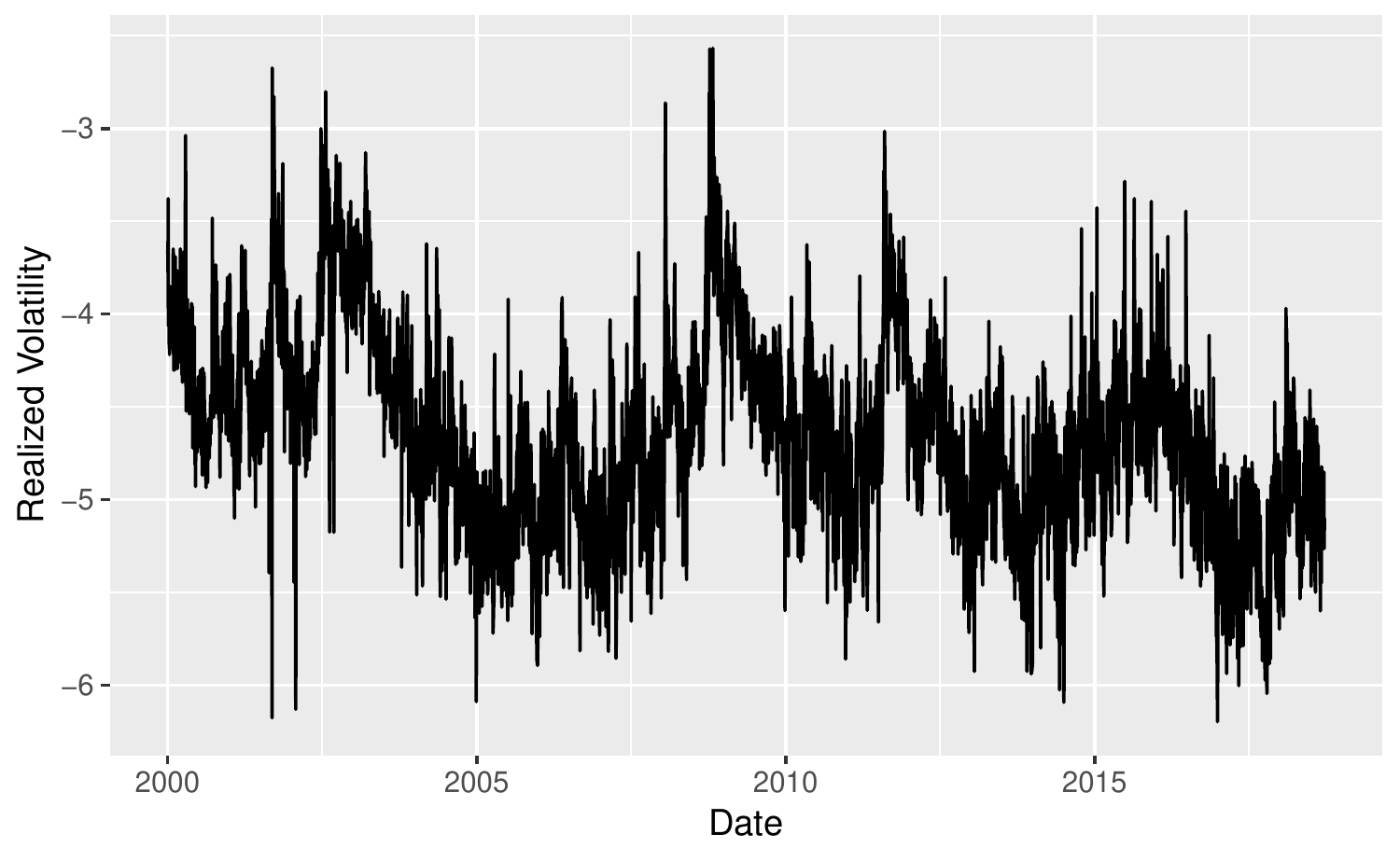}
		\caption{timeplot}
		\label{fig:timeplotRV_DAX}
	\end{subfigure}
	\begin{subfigure}{0.45\textwidth}
		\centering
		\includegraphics[width=1\linewidth]{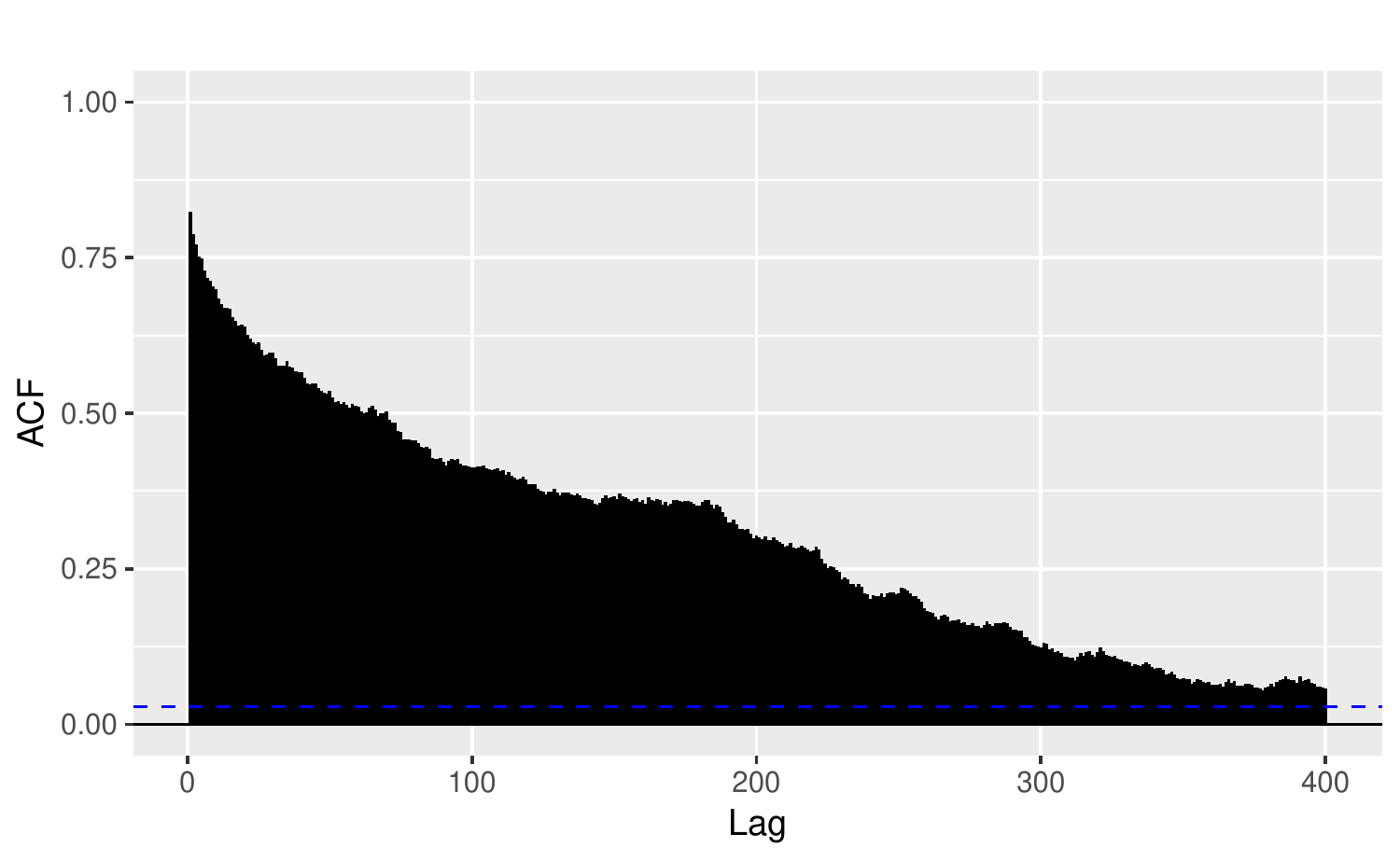}
		\caption{autocorrelogram}
		\label{fig:correlogramRV_DAX}
	\end{subfigure}
	\caption{plots for daily DAX realized volatility $\log(\widehat{\sigma}_t)$ from 3.01.2000 to 20.09.2018}
	\label{fig:plotRV_DAX}
\end{figure}

We also changed the country and used the respective series for the DAX, again from January 2000 to September 2018. It is plotted in Figure \ref{fig:plotRV_DAX} together with its autocorrelation function, which looks very similar to the previous one. The results as shown in Table \ref{RV3} do not change substantially, too, with the only discrepancy that now for the smaller bandwidths and for the larger horizons some performance gains of the methods with estimated $d$ over the fixed-$d$ method occur.

\begin{table}[h]
	\centering
	\caption{MSEs relative to FI($0.5$) from POOS forecasting experiments for daily realized volatility $\log(\widehat{\sigma}_t)$ of the DAX, from 3.01.2000 to 20.09.2018, rolling window of size 500}
	\label{RV3}
	\begin{adjustbox}{width=0.8\textwidth}
		\begin{tabular}{@{}lllllllll@{}}
			\toprule
			horizon & FI ($T^{0.5}$) &  FI ($T^{0.65}$) &  FI ($T^{0.8}$)  & FI ($1$)   & LAR &  ES & Mean & HAR \\ \midrule
			$h=1$ &\textit{0.997}&\textbf{0.996}&1.001&1.019&1.014&1.011&3.405&\textit{0.997}\\
			$h=3$ &\textit{0.992}&\textbf{0.989}&0.998&1.021&1.022&1.022&2.693&0.996\\
			$h=5$ &\textit{0.991}&\textbf{0.987}&0.995&1.039&1.04&1.044&2.423&0.999\\
			$h=10$ &\textit{0.989}&\textbf{0.982}&0.996&1.047&1.063&1.065&2.058&1.014\\
			$h=20$ &\textit{0.98}&\textbf{0.965}&0.993&1.05&1.096&1.073&1.75&1.022\\
			$h=40$ &\textit{0.96}&\textbf{0.946}&0.993&1.043&1.135&1.06&1.519&1.045\\
			$h=80$ &\textit{0.939}&\textbf{0.937}&0.993&1.057&1.135&1.078&1.325&1.058\\
			\bottomrule
		\end{tabular}
	\end{adjustbox}
\end{table}

\subsection{Experiments on the optimal estimation window size}

An important and often overlooked specification issue when setting up a forecasting method is the size of the estimation window, see e.g.\ \cite{pesaran2007}. Should all available information on the past of the series of interest be used to decrease the variance of the estimators or should only the more recent past be included to robustify the procedure against structural change? Certainly, the answer to this question depends on the characteristics of the individual variables. But as it is often implicitly assumed that a longer stretch of past data leads to better results, we performed experiments on the optimal window size on the inflation and realized volatility series used already in this paper to check this assumption. At the same time, the experiments in this section serve as a robustness check for the choice of the window size in the first two parts of this section. Moreover, the effects of using an expanding instead of a rolling estimation window are analyzed.

We again employed pseudo-out-of-sample forecasting for different horizons and now also for different sizes of the rolling estimation window and for an expanding window. This enables us to compare the accuracy of forecasts obtained by using rolling windows of different sizes and an expanding window. To ensure a better comparability, we let the methods all produce forecasts for the same time period, which means that for the shorter window sizes the remaining first parts of the series, which are needed for the estimation with the longer sizes, are not used.  

\begin{table}[h]
	\centering
	\caption{MSEs from POOS experiments for different rolling window sizes and for an expanding window with initial size 360 for monthly year-on-year US CPI inflation, January 1948 to November 2017, the first forecast is for January 1988 for all sizes, forecast horizon $h=1$}
	\label{WS1}
	\begin{adjustbox}{width=0.9\textwidth}
		\begin{tabular}{@{}lllllllll@{}}
			\toprule
			window size & FI ($T^{0.5}$) &  FI ($T^{0.65}$) &  FI ($T^{0.8}$)  & FI ($0.5$) & FI ($1$)   & LAR & ES & Mean  \\ \midrule
			$60$ &\textbf{0.1322}&\textit{0.1333}&0.1338&0.1337&0.1343&0.1536&0.1515&4.3899\\
			$120$
			&0.1051&0.1032&\textbf{0.1019}&0.1039&\textit{0.1031}&0.1072&0.1505&5.5664\\
			$180$ &0.1036&0.1044&\textit{0.1033}&\textbf{0.1028}&0.1055&0.1049&0.1527&6.815\\
			$240$ &\textbf{0.1029}&0.1035&0.1047&0.1034&0.105&\textit{0.1032}&0.1558&8.2456\\
			$360$ &0.106&0.1048&\textit{0.1047}&\textbf{0.104}&\textit{0.1047}&0.1053&0.1481&9.4418\\
			$expanding$ &\textit{0.1055}&0.1067&0.1066&\textbf{0.105}&0.1064&0.1061&0.1445&8.0776\\
			\bottomrule
		\end{tabular}
	\end{adjustbox}
\end{table}

Tables \ref{WS1} and \ref{WS2} contain the results for the US inflation series from the beginning of this section for the forecasting horizons $h=1$ and $h=6$ respectively and for rolling windows of sizes varying from 60 to 360 and an expanding window of initial size 360. For the other inflation series and other horizons the results are very similar and so we do not report them here. We now report absolute MSFEs to enable a comparison over the rows and hence a column for the FI($0.5$) method is added. There are basically two ways of reading the tables: Firstly, the focus of interest can be the best forecasting performance over all methods and window sizes. Then only the results of the best method for each window size needs to be considered, which amounts to a comparison of the numbers in bold face. Secondly, when interested in the dependence of the performance of a single method on the window size one should stay in the respective column. Taking the first perspective, the forecasting performance increases strongly from a window size of 60 to a size of 120. From then on, the MSE does not change much anymore, but seems to rise slightly for the largest size of 360 and for the expanding window. Looking at the individual methods, the same pattern emerges for the FI-based methods and the autoregression. Hence, for the monthly inflation series a rolling window with a medium estimation window size between 120 and 240 seems to be a good choice, balancing out an increased estimation uncertainty for too small sizes and the vulnerability to gradual structural change for too large sizes optimally.  

\begin{table}[h]
	\centering
	\caption{MSEs from POOS experiments for different rolling window sizes and for an expanding window with initial size 360 for monthly year-on-year US CPI inflation, January 1948 to November 2017, the first forecast is for June 1988 for all sizes, forecast horizon $h=6$}
	\label{WS2}
	\begin{adjustbox}{width=0.9\textwidth}
		\begin{tabular}{@{}lllllllll@{}}
			\toprule
			window size & FI ($T^{0.5}$) &  FI ($T^{0.65}$) &  FI ($T^{0.8}$)  & FI ($0.5$) & FI ($1$)   & LAR & ES & Mean   \\ \midrule
			$60$ &\textit{1.6622}&\textbf{1.6441}&1.7114&1.8034&1.707&3.1693&1.9453&5.1008\\
			$120$ &1.6596&\textit{1.6114}&\textbf{1.599}&1.6434&1.6472&1.8125&1.8672&6.0981\\
			$180$ &\textit{1.6103}&1.6421&1.6676&\textbf{1.6073}&1.6701&1.7181&1.8615&7.3252\\
			$240$ &1.653&\textbf{1.5387}&1.5603&\textit{1.5417}&1.5441&1.5881&1.8457&8.6649\\
			$360$ &1.7027&1.6369&1.6364&\textbf{1.6018}&1.64&\textit{1.6064}&1.8826&9.697\\
			$expanding$ &\textit{1.6351}&1.7755&1.7525&\textbf{1.6172}&1.7143&1.665&1.8027&8.2271\\
			\bottomrule
		\end{tabular}
	\end{adjustbox}
\end{table}

For the realized volatility series from above we did the same analysis for rolling window sizes from 125 to 1000 and an expanding window of size 1000. We report the results for forecast horizons $h=1$ and $h=5$ for the initial series from above in Tables \ref{WS3} and \ref{WS4}, the results for other horizons and window sizes are very similar as well. Again, picking a too small window size of 125 leads to a drop in performance. From the window size 250 on the MSEs stay almost constant for increasing window sizes of the rolling window up to 1000 and are almost identical to the results when using an expanding window. Thus, for the realized volatility series it does hardly matter which window size is chosen as long as it is not to small. 

\begin{table}[h]
	\centering
	\caption{MSEs from POOS experiments for different rolling window sizes and for an expanding window with initial size 1000 for daily realized volatility $\log(\widehat{\sigma}_t)$ of the S\&P 500, from 3.01.2000 to 20.09.2018, the first forecast is for the 1001st observation for all sizes, forecast horizon $h=1$}
	\label{WS3}
	\begin{adjustbox}{width=0.9\textwidth}
		\begin{tabular}{@{}lllllllll@{}}
			\toprule
			window size & FI ($T^{0.5}$) &  FI ($T^{0.65}$) &  FI ($T^{0.8}$)  & FI ($0.5$) & FI ($1$)   & LAR & ES & Mean   \\ \midrule
			$125$ &0.0959&\textit{0.0949}&\textbf{0.0945}&0.0951&0.1003&0.098&0.0951&0.1969\\
			$250$ &0.0929&\textit{0.0921}&\textbf{0.0915}&0.0924&0.097&0.094&0.0937&0.2283\\
			$500$
			&0.092&\textit{0.0916}&\textbf{0.0912}&0.0917&0.0944&0.0934&0.0938&0.2807\\
			$750$&0.0919&\textit{0.0917}&\textbf{0.0915}&0.0918&0.0941&0.0934&0.0942&0.3063\\
			$1000$&0.092&\textbf{0.0917}&\textbf{0.0917}&0.0919&0.0933&0.0935&0.0943&0.3266\\
			$expanding$ &0.0919&0.0919&\textbf{0.0917}&\textit{0.0918}&0.0929&0.0923&0.0947&0.369\\
			\bottomrule
		\end{tabular}
	\end{adjustbox}
\end{table}

\begin{table}[h]
	\centering
	\caption{MSEs from POOS experiments for different rolling window sizes and for an expanding window with initial size 1000 for daily realized volatility $\log(\widehat{\sigma}_t)$ of the S\&P 500, from 3.01.2000 to 20.09.2018, the first forecast is for the 1005th observation for all sizes, forecast horizon $h=5$}
	\label{WS4}
	\begin{adjustbox}{width=0.9\textwidth}
		\begin{tabular}{@{}llllllllll@{}}
			\toprule
			window size & HW & FI ($T^{0.5}$) &  FI ($T^{0.65}$) &  FI ($T^{0.8}$)  & FI ($0.5$) & FI ($1$)   & LAR & ES & Mean \\ \midrule
			$125$ &0.1504&0.1494&0.1458&\textbf{0.1453}&\textbf{0.1453}&0.1623&0.1562&0.1585&0.2064\\
			$250$ &0.1485&0.1432&0.1418&\textit{0.1416}&\textbf{0.1413}&0.1599&0.151&0.1571&0.2337\\
			$500$ &0.1488&0.1415&\textit{0.141}&\textbf{0.1408}&0.1416&0.1513&0.1497&0.1565&0.2847\\
			$750$ &0.1481&0.1425&\textbf{0.1408}&\textit{0.1409}&0.1415&0.1504&0.1493&0.1564&0.309\\
			$1000$ &0.1472&0.1423&\textbf{0.1411}&\textit{0.1412}&\textit{0.1412}&0.1482&0.1485&0.1558&0.3289\\
			expanding &0.1442&0.1414&\textit{0.1411}&\textbf{0.1409}&0.1413&0.146&0.1438&0.1538&0.37\\
			\bottomrule
		\end{tabular}
	\end{adjustbox}
\end{table}

%SSSSSSSSSSSSSSSSSSSSSSSSSSSSSSSSSSSSSSSSSSSSSSSSSSSSSSSSSSSSSSSSSSSS
\section{Simulation evidence}

To analyze if the empirical findings from the previous section still hold when true FI processes are forecasted and if some of these findings can be better explained by systematic variations in the data generating process, we conducted a simulation study. We simulated from the FI process given in equation \ref{FI_typeII} with different values of the long memory parameter $d$ and the three different short memory components $\{x_t\}$ already used in sections 3 and 4. For each instance we generated 1000 series of size $T+h$, held back the last $h$ observations and used the first $T$ observations to generate the forecasts for period $T+h$. We chose $T \in \{60,300,1500\}$ and $h \in \{1,3,12\}$. As in the previous section we present the MSEs relative to FI(0.5) and the smallest and second smallest MSEs in each row of the tables are marked in bold face and italics respectively.

Table \ref{simu_1} contains the results for FI(0.4) processes. The FI(0.5) method again almost uniformly outperforms the long autoregression and the exponential smoothing method. This confirms the empirical results and further makes a case for the use of true long memory models when forecasting strongly persistent time series. Additionally, as observed with the empirical POOS experiments, the predictive performance of both LAR and ES gets worse relative to FI(0.5) as the forecasting horizon increases. Unsurprisingly, adequately accounting for long memory seems to get more critical with increasing forecasting horizon. While the dynamics over a short horizon may be approximated by only a short memory component, over longer horizons the long memory component increasingly dominates the forecastable part of the process as the influence of the short memory component dies out much faster. Furthermore, as in the empirical experiments, the long autoregression clearly shows the best performance from the methods that do not employ a long memory model.

When comparing the performance of the different FI-based methods, the ones using the local Whittle estimator with different bandwidth choices to estimate the long memory parameter do not outperform the one using an a priori fixed $d=0.5$ (although this is not the true $d$), not even for the largest sample size $T=1500$. As expected, the performance of the FI($T^{\alpha}$) methods varies with the bandwidth and for more persistent $\{x_t\}$ larger bandwidths lead to better results. The smaller the sample size and the larger the persistence of $\{x_t\}$ gets, the worse do the FI($T^{\alpha}$) methods come off against FI($0.5$). This can be explained by the lack of accuracy and the bias respectively that plague the local Whittle estimator under these circumstances. Regarding the influence of the forecasting horizon, exactly under these circumstances, where the estimation of $d$ is difficult, a similar phenomenon occurs as for LAR and ES: The relative performance compared to FI(0.5) decreases with $h$, sometimes leading to very large losses for $h=12$. This behavior is problematic as these conditions often arise in practice: Longer forecasting horizons are naturally of interest under long memory, small sizes of the estimation sample will rather be the rule than the exception, especially given the analysis at the end of the last section concerning optimal estimation window sizes, and substantive short memory dynamics may well arise. As discussed above, the long memory component becomes more and more important the longer the forecasting horizon gets. Thus, as the difficulty of disentangling long and short memory dynamics and estimating the long memory component often leads to a very wrong specification of the long memory component, the predictive performance of the FI($T^{\alpha})$ methods suffers. Our experiments suggest that it works better to accept a wrong value of $d$ a priori and choose it ad hoc so that it lies at least not very far away from the dynamics usually observed in economic time series. Together with a short memory component which may somehow adapt to this, a good and robust forecasting performance can be achieved.

\begin{table}[h]
	\centering
	\caption{MSEs relative to FI($0.5$) for different forecasting methods for simulated series of size $T$ and a forecast horizon $h$, the series are generated by FI($0.4$) processes with short memory component $x_t$, 1000 repetitions}
	\label{simu_1}
	\begin{adjustbox}{width=0.8\textwidth}
		\begin{tabular}{@{}llllllllll@{}}
			\toprule
			$x_t$  & $T$  & $h$ & FI ($T^{0.5}$) &  FI ($T^{0.65}$) &  FI ($T^{0.8}$)  & FI ($1$) & LAR   & ES & Mean  \\ \midrule 
			iid	 & $60$ & $1$ &1.011&\textit{0.974}&\textbf{0.964}&1.117&1.049&\textit{0.974}&1.243\\	
			& &  $3$ &1.043&\textit{1.019}&\textbf{0.995}&1.236&1.072&1.123&1.041\\
			& & $12$ &1.058&1.014&\textit{0.997}&1.281&1.093&1.221&\textbf{0.985}\\
			& $300$ & $1$ &1.006&\textit{0.999}&\textbf{0.993}&1.036&1.028&1.002&1.567\\
			& & $3$ &1.042&\textit{1.017}&\textbf{1.008}&1.118&1.065&1.099&1.329\\
			& & $12$  &1.006&\textit{0.991}&\textbf{0.987}&1.197&1.049&1.24&1.093\\
			& $1500$ & $1$ &1.002&\textit{0.999}&\textbf{0.995}&1.01&1.013&1.069&1.754\\
			& &  $3$ &1.004&\textit{0.998}&\textbf{0.995}&1.061&1.014&1.079&1.376\\
			& & $12$ &\textit{0.988}&0.99&\textbf{0.986}&1.121&1.022&1.219&1.151\\
			\\
			AR(1)	 & $60$ & $1$ 	&\textit{1.001}&1.017&1.018&1.056&1.023&\textbf{0.98}&3.096\\
			& &  $3$ &\textbf{1.025}&1.049&1.053&1.186&\textit{1.037}&1.147&1.285\\
			& & $12$ &1.097&1.079&1.098&1.326&\textit{1.069}&1.327&\textbf{1.027}\\
			& $300$ & $1$ &\textbf{1.002}&\textit{1.003}&1.012&1.023&1.015&1.037&3.897\\
			& & $3$ &\textit{1.011}&\textbf{1.007}&1.036&1.105&1.03&1.217&1.939\\
			& & $12$  &\textit{1.019}&\textbf{1.017}&1.075&1.278&1.051&1.434&1.151\\
			& $1500$ & $1$ &\textit{1.002}&\textbf{0.999}&\textit{1.002}&1.028&1.013&1.071&5.345\\
			& &  $3$ &\textbf{0.99}&\textbf{0.99}&1.016&1.063&1.017&1.223&1.898\\
			& & $12$ &\textbf{0.993}&\textit{0.996}&1.035&1.173&1.035&1.466&1.298\\	
			\\
			MA(9)  & $60$ & $1$ &1.052&1.053&1.054&\textbf{1.026}&\textit{1.043}&1.064&16.356\\	
			& &  $3$ &\textit{1.099}&1.103&1.107&\textbf{1.051}&1.11&1.137&3.369\\
			& & $12$ &1.511&1.579&1.383&\textit{1.239}&1.359&1.638&\textbf{1.036}\\
			& $300$ & $1$ &\textbf{1.002}&1.013&1.032&1.018&\textit{1.011}&1.039&22.784\\
			& & $3$ &\textit{1.012}&1.051&1.089&1.045&\textbf{1.01}&1.121&4.666\\
			& & $12$  &\textbf{1.027}&1.235&1.485&1.282&\textit{1.033}&1.744&1.327\\
			& $1500$ & $1$ &0.998&\textit{0.995}&1.01&1.001&\textbf{0.989}&1.031&27.057\\
			& &  $3$ &\textit{1.001}&1.002&1.042&1.025&\textbf{0.999}&1.171&5.704\\
			& & $12$ &\textbf{1.002}&\textit{1.018}&1.231&1.164&1.024&1.755&1.433\\
			\bottomrule
		\end{tabular}
	\end{adjustbox}
\end{table}

Table \ref{simu_2} presents the results for simulated nonstationary long memory processes, namely FI(0.7) processes. As expected, as the true value is further away from 0.5 now, the other methods gain a little ground relative to FI($0.5$). But still no large gains are possible by estimating $d$, the improvements under the best circumstances and the best bandwidth choice for FI($T^{\alpha}$) never reach 5 percent. Otherwise, no substantive changes occur compared to Table \ref{simu_1}. Exponential smoothing performs better than above, but is still not as good as the long autoregression. 

\begin{table}[h]
	\centering
	\caption{MSEs relative to FI($0.5$) for different forecasting methods for simulated series of size $T$ and a forecast horizon $h$, the series are generated by FI($0.7$) processes with short memory component $x_t$, 1000 repetitions}
	\label{simu_2}
	\begin{adjustbox}{width=0.8\textwidth}
		\begin{tabular}{@{}llllllllll@{}}
			\toprule
			$x_t$ & $T$  & $h$ & FI ($T^{0.5}$) &  FI ($T^{0.65}$) &  FI ($T^{0.8}$)  & FI ($1$) & LAR   & ES & Mean  \\ \midrule 
			iid & $60$ & $1$ &1.008&0.982&\textit{0.974}&1.06&1.058&\textbf{0.943}&3.488\\	
			& &  $3$ &1.008&\textit{0.992}&\textbf{0.965}&1.06&1.04&1.024&1.787\\
			& & $12$ &1.062&\textit{0.981}&\textbf{0.959}&1.05&1.208&1.038&1.165\\
			& $300$ & $1$ &0.989&0.989&\textit{0.975}&1.001&1.005&\textbf{0.964}&6.663\\
			& & $3$ &1.001&\textit{0.976}&\textbf{0.968}&1.024&1.035&1.022&4.165\\
			& & $12$  &0.994&\textit{0.969}&\textbf{0.954}&1.08&1.095&1.124&2.121\\
			& $1500$ & $1$ &0.992&\textit{0.989}&\textbf{0.987}&1.002&0.999&1.031&14.521\\
			& &  $3$ &0.985&\textit{0.98}&\textbf{0.979}&1.01&1.005&1.042&7.857\\
			& & $12$ &0.999&\textit{0.997}&\textbf{0.99}&1.073&1.053&1.133&3.866\\
			\\
			AR(1)	& $60$ & $1$ &\textit{1.016}&1.043&1.03&1.038&1.096&\textbf{0.997}&11.316\\
			& & $3$ &1.049&1.04&\textit{1.036}&1.04&1.106&\textbf{0.986}&3.062\\
			& & $12$ &1.131&\textbf{1.078}&\textit{1.085}&1.089&1.193&1.162&1.331\\
			& $300$ & $1$ &\textit{0.988}&0.989&0.994&0.998&0.994&\textbf{0.979}&25.032\\
			& & $3$ &\textbf{0.986}&\textit{0.988}&1.015&1.021&1.012&1.063&7.634\\
			& & $12$  &\textit{0.979}&\textbf{0.973}&1.023&1.046&1.048&1.09&2.217\\
			& $1500$  & $1$ & \textit{0.993}&\textbf{0.99}&\textit{0.993}&0.998&1&1.019&58.401\\
			& & $3$ &\textit{0.989}&\textbf{0.987}&1.012&1.022&1.009&1.082&13.297\\
			& & $12$ &\textbf{0.977}&\textbf{0.977}&1.03&1.063&1.026&1.145&4.757\\
			\\
			MA(9)  & $60$ & $1$ & 1.046&1.049&1.041&\textbf{1.022}&1.058&\textit{1.034}&76.782\\
			& &  $3$ &1.051&1.038&\textit{1.033}&\textbf{0.983}&1.102&1.047&10.086\\
			& & $12$ &1.327&1.313&\textit{1.116}&\textbf{1.019}&1.402&1.558&1.405\\
			& $300$ & $1$ &\textbf{0.999}&1.011&1.017&1.014&1.027&\textit{1.001}&186.109\\
			& & $3$ &\textbf{1.008}&1.034&1.047&\textit{1.01}&1.017&1.019&24.082\\
			& & $12$  &\textbf{1.027}&1.179&1.274&1.071&\textit{1.06}&1.244&3.309\\
			& $1500$ & $1$ &1.006&1.003&1.01&\textit{1.002}&\textbf{0.998}&1.005&391.329\\
			& &  $3$ &\textbf{0.998}&1&1.024&1.003&\textit{0.999}&1.046&47.794\\
			& & $12$ &\textbf{0.987}&1.002&1.17&1.022&\textit{0.993}&1.226&5.836\\
			\bottomrule
		\end{tabular}
	\end{adjustbox}
\end{table}

The results from the simulation study are perfectly in line with the ones from the ones from the empirical section and additionally help to better understand them: For the inflation series we observed that the methods with estimated $d$ showed a worse performance than the method with the fixed $d=0.5$ and the performance dropped over the forecasting horizon, usually with very large losses for large horizons. As we observed an analogous behavior for series with more persistent short memory components, the inflation series seem to fit in this category. For realized volatility, the two methods show a very similar performance, with perhaps slight advantages for the methods with estimated $d$. Thus, the realized volatility series series do not seem to have very persistent short memory components. Furthermore, the estimation window sizes used for the realized volatility series are longer due to the higher frequency of the financial series. 

We conclude from the forecasting experiments with empirical and simulated series presented in the last two sections that explicitly accounting for long memory by using FI models leads to improved results. When using a FI model, fixing the parameter to $d=0.5$ a priori offers a generally well-performing and simple method, which is hard to beat and could serve as a benchmark for forecasting under long memory. In some situations, slight improvements over this method are possible by using a semiparametric estimator for $d$, having to accept a higher downside risk.

%SSSSSSSSSSSSSSSSSSSSSSSSSSSSSSSSSSSSSSSSSSSSSSSSSSSSSSSSSSSSSSSSSSSSS
\section{Concluding remarks}

In this paper we are concerned with the issue of forecasting strongly persistent time series. By pseudo-out-of-sample forecasting experiments on inflation and realized volatility time series and by simulations we find clear evidence that accounting explicitly for long memory by using a fractionally integrated model improves predictive performance in comparison to classical short memory competitors. We are able to explain the former mixed evidence on this issue and to present a simple and successful method for forecasting by proposing the use of a fixed-parameter FI model with $d=0.5$, which overcomes the problem of the high uncertainty associated with the estimation of the long memory parameter and is nevertheless able to capture decisive long memory features.

While our results hence suggest that FI models are a useful tool in forecasting strongly persistent time series, which are widespread in economics, finance and many other fields, there seems not much too be gained by estimating $d$, i.e.\ by trying to determine the exact strength of the long memory. Further support for this is provided by our analysis of the optimal estimation window size, where we find that medium window sizes provide an optimal balance between high estimation variances and gradual structural change. 

To construct proper forecasting methods from FI models, the estimation of the mean and of the long memory parameter (at least if one wants to use models with an estimated parameter or wants to study the effects of the estimation) are crucial. We investigate both and our results are of general interest outside the realm of forecasting as well. We compare the arithmetic mean and Shimotsu's and Robinson's mean estimators by a theoretical analysis, providing asymptotic theory for the latter for the first time, and by a simulation study to find that the Robinson estimator can yield efficiency gains and shows a good performance in practice. Regarding the estimation of $d$, we compare the Whittle estimator plus AIC, the global semiparametric Whittle and the local Whittle estimator to examine if parametric estimators plus model selection step, global or local semiparametric estimators yield more accurate results. Previous literature has already identified the members of the Whittle family as favourable. As the parametric and the global semiparametric estimator incur drastic losses compared to the local Whittle in our simulations, which we can trace back to the very bad performance of the Whittle estimator under even small amounts of misspecification, we conclude that they should be used with care and that local semiparametric estimators should usually be a better choice. 

Not to go beyond the scope of this paper, we limit our analysis to a univariate framework, but we believe that many of our fundamental findings will continue to hold and be helpful in multivariate settings. Furthermore, for the fixed-$d$ FI processes we picked the value $d=0.5$ at the beginning of our analysis and do not report results for alternative choices, withstanding the statistician's immanent desire to optimize parameter values (after just having overcome the estimation problem). Nevertheless, it is certainly possible to go in this direction and to develop methods for the selection of the value of $d$ based on predictive performance. \cite{Baillie_etal14b} for example do a similar thing and pick the bandwith for semiparametric estimation of $d$ by cross-validation. Further, it seems natural to combine forecasts from different FI specifications in order to robustify against model misspecification, see e.g.\ \cite{timmermann2006} on forecast combination. In all of these directions there is certainly room for future research.

\section*{Appendix}

{\sc Proof of Proposition 1} The variance  is given by
\[
\var(\widehat \mu (d)) = \var \left( \sum_{t=1}^T r_t x_t \right)  \left( \sum_{t=1}^T r_t^2 \right)^{-2} ,
\]
where $r_t = \sum_{j=0}^{t-1} \pi_j (d)$. In a first step, we deal with the denominator  of this variance, in a second step we will turn to the numerator.

Step 1) From \citet[Lemma 5.1]{Hassler19} we have $r_t = \pi_{t-1} (d-1)$, which implies $r_t \sim t^{-d}/\Gamma (1-d)$. From \citet[Prop. 2.3.1]{Giraitisetal12} we have $\sum_{t=1}^T t^{-2d} \sim T^{1-2d} / (1-2d)$. Now, define $s_T=\sum_{t=1}^T r_t^2$ and $\sigma_T :=\sum_{t=1}^T t^{-2d}$, where the latter constitutes a  strictly monotone and divergent sequence. Note that $(s_{T}-s_{T-1})/(\sigma_{T}-\sigma_{T-1}) \to 1 / \Gamma^2 (1-d)$. Hence, we can use the so-called Stolz-Ces\`{a}ro Theorem, see \citet[pp. 76, 77]{Knopp90}, and conclude that $s_T \sim \sigma_T / \Gamma^2 (1-d)$. This provides
\[
\sum_{t=1}^T r_t^2 \ \sim \ \frac{T^{1-2d}}{(1-2d) \Gamma^2 (1-d)} \, .
\]

Step 2) The numerator becomes
\[
\var \left( \sum_{t=1}^T r_t x_t \right)  = \gamma_x (0) \sum_{t=1}^T r_t^2 + 2 \sum_{h=1}^{T-1} \gamma_x (h) \sum_{t=1}^{T-h} r_t r_{t+h}.
\]
With $r_{t+h} = r_{t} + \sum_{j=t}^{t+h-1} \pi_j (d)$ we obtain
\[
\frac{ \sum_{t=1}^{T-h} r_t r_{t+h}}{\sum_{t=1}^T r_t^2} = 1- \frac{\sum_{t=T-h+1}^T r_t^2 - \sum_{t=1}^{T-h} r_t \sum_{j=t}^{t+h-1} \pi_j (d)}{\sum_{t=1}^T r_t^2} = 1 - \frac{b_h}{b_T} \, ,
\]
where ${b_h}$ and ${b_T}$ are defined implicitly. Note that $\pi_0 (d)=1$ and  $\pi_j (d)$ is negative and monotonically increasing for $j > 0$.  Consequently, the sequence $\{b_h\}$ is positive and monotonically increasing, while $b_T$ diverges according to the previous step. Hence,
\[
\sum_{h=1}^{T-1} \frac{b_h \gamma_x (h)}{b_T} \ \to \ 0 \quad \mbox{as } T \to \infty \, ,
\]
by what is sometimes called Kronecker's Lemma. It follows that
\[
\frac{\var \left( \sum_{t=1}^T r_t x_t \right) }{\sum_{t=1}^T r_t^2} \ \to \ \gamma_x (0) + 2 \sum_{h=1}^{\infty} \gamma_x (h) = \omega_x^2 \, .
\]
Hence, the proof is complete.

\bibliography{forecasting_LM}
\bibliographystyle{chicago}

\end{document}